\documentclass{emulateapj} 

\bibpunct{(}{)}{;}{a}{}{,} 
 
\slugcomment{Accepted Dec 6, 2006} 
 
\shorttitle{Abundance Analysis of HE~1300+0157}  
 
\shortauthors{Frebel et al.} 
 
\begin{document} 
\title{Chemical Abundance Analysis of the \\Extremely Metal-Poor Star 
HE~1300$+$0157\altaffilmark{*}}

\author{Anna Frebel\altaffilmark{1},  
John E. Norris\altaffilmark{1},  
Wako Aoki \altaffilmark{2},  
Satoshi Honda \altaffilmark{2},  
Michael S. Bessell \altaffilmark{1},\\ 
Masahide Takada-Hidai \altaffilmark{3}, 
Timothy C. Beers \altaffilmark{4},  
and Norbert Christlieb\altaffilmark{5,6}}

\altaffiltext{*}{Based on data collected at the Subaru Telescope, which is 
operated by the National Astronomical Observatory of Japan.} 
\altaffiltext{1}{Research School of Astronomy \& Astrophysics, The Australian 
National University, Cotter Road, Weston, ACT 2611, Australia} 
\altaffiltext{2}{National Astronomical Observatory of Japan, 2-1-21 Osawa, 
Mitaka, Tokyo, 181-8588 Japan}  
\altaffiltext{3}{Liberal Arts Education Center, Tokai University, 
1117 Kitakaname, Hiratsuka-shi, Kanagawa 259-1292, Japan} 
\altaffiltext{4}{Department of Physics \& Astronomy, CSCE: Center for the Study of Cosmic Evolution, 
and JINA: Joint Institute for Nuclear Astrophysics, Michigan State University, East Lansing, MI 48824-1116, 
U.S.A} 
 
\altaffiltext{5}{Hamburger Sternwarte, 
Gojenbergsweg 112, 21029 Hamburg, Germany}  
 
\altaffiltext{6}{Department of 
Astronomy and Space Physics, Uppsala University, Box 515, 751-20 Uppsala, 
Sweden} 

\begin{abstract} 
We present a detailed chemical abundance analysis of HE~1300$+$0157, a
subgiant with \mbox{$\mbox{[Fe/H]}=-3.9$}. From a high-resolution, high-$S/N$
Subaru/HDS spectrum we find the star to be enriched in C ($\mbox{[C/Fe]}_{\rm
1D}\sim +1.4$) and O ($\mbox{[O/Fe]}_{\rm 1D}\sim +1.8$). With the exception
of these species, HE~1300$+$0157 exhibits an elemental abundance pattern
similar to that found in many other very and extremely metal-poor stars. The
Li abundance is lower than the Spite-plateau value, in agreement with
expectation for its evolutionary status. Of particular interest, no
neutron-capture elements are detected in HE~1300$+$0157. This type of
abundance pattern has been found by recent studies in several other metal-poor
giants. We suggest that HE~1300$+$0157 is an unevolved example of this group
of stars, which exhibit high C abundances together with low (or absent)
abundances of neutron-capture elements (CEMP-no). Several potential enrichment
scenarios are presented.  The non-detection of neutron-capture elements
including Sr, Ba, and Pb suggests that the carbon excess observed in
HE~1300$+$0157 is not due to mass transfer across a binary system. Such a
scenario is applied to carbon-rich objects with excesses of s-process
elements. The normal observed Li abundance supports this interpretation. Most
likely, the high levels of C and O were produced prior to the birth of this
star. It remains unclear whether a single hypernova is responsible for its
overall chemical pattern, or whether one requires a superposition of yields
from a massive Population\,III object and a normal Type II SN. These scenarios
provide important information on the C production in the early Universe, and
on the formation of C-rich stars in the early Galaxy.
\end{abstract} 
 
 
 
\section{Introduction} 
 
 
The study of the most metal-poor stars offers a unique opportunity to learn
about conditions in the early Galaxy, the first stars and supernovae (SNe),
and the beginning of chemical enrichment in the Universe. This information is,
however, difficult and challenging to obtain.  For example, only 17
stars\footnote{This number depends on the adopted temperature scales of the
individual analyses.} with high resolution, high-$S/N$ analyses are currently
known with metallicities $\mbox{[Fe/H]}\lesssim-3.5$. Ten of these have
$\mbox{[Fe/H]}<-3.7$; the number shrinks to two below
$\mbox{[Fe/H]}\sim-4.1$. Surprisingly, the stars exhibit a variety of
different chemical abundance patterns. It is thus difficult to derive
definitive conclusions about the chemical conditions of their birth clouds, or
which processes enriched the objects. What is apparent from this diversity is
that, despite their similar (low) iron abundances, these stars do not all
share a common origin, and that nucleosynthesis processes contributed to the
chemical enrichment of the early Galaxy in many different ways. It naturally
follows that abundance trends at the lowest metallicities (below, say,
$\mbox{[Fe/H]}\sim-3.0$) are not well established. These in turn are important
ingredients for Galactic Chemical Evolution models (e.g.,
\citealt{chiappini_1999,karlsson2005}) or detailed modeling of the first
supernovae (e.g., \citealt{heger2002,UmedaNomotoNature}). Clearly, given the
small number of the lowest-metallicity objects known at present, a larger
sample is needed to derive more definite conclusions about our Galactic past,
both observationally and theoretically.
 
A large sample of very metal-poor stars also allows investigation of the shape
of the metallicity distribution function of the stellar halo.  Some
interesting details are beginning to emerge.  For instance, there seems to
exist a gap in the interval $-5\lesssim\mbox{[Fe/H]}\lesssim-4$, or at least
no star has yet been found in this particular range. Some authors (e.g.,
\citealt{shigeyama, karlsson2006}) have suggested that this may be the result
of an underlying astrophysical process, rather than just small-number
statistics. The discoveries of further stars in this metallicity range will
test these ideas.
 
Using the High Dispersion Spectrograph (HDS) at the Japanese 8\,m Subaru
telescope on Mauna Kea, Hawai'i, we have conducted an extensive program over
several semesters to discover more ultra-metal-poor (UMP) stars with
$\mbox{[Fe/H]}\lesssim-3.5$. The candidates of this project are selected from
the HK survey \citep{beers_stromlo_symp}, as well as from the faint
\citep{Christlieb:2003} and bright \citep{frebel_bmps} objects of the
Hamburg/ESO survey (HES).  Of the stars with $\mbox{[Fe/H]}\lesssim-3.0$ that
have been identified in our campaign, 14 are already analyzed
\citep{Aoki_IAU05}. The most well-known star of this group is HE~1327$-$2326,
with $\mbox{[Fe/H]}=-5.4$ \citep{HE1327_Nature, Aokihe1327}. Further results
of the UMP project will soon be reported (W. Aoki et al. 2006, in
preparation).
 
In this paper we present HE~1300+0157, a star selected from the faint HES. The
medium-resolution follow-up spectrum was obtained with the 6dF instrument at
the UK Schmidt telescope, located at Siding Spring Observatory,
Australia. Based on the low metallicity measured from those data, HE~1300+0157
was selected for snapshot\footnote{Snapshot observation refers to a
$\sim20$\,min exposure with $R\sim20,000$ that is sufficient for a basic
high-resolution abundance analysis.} spectroscopic observation with UVES
at the VLT, as part of the HERES project (see \citealt{heresI} for further
details). From the UVES spectrum, \citet{heresII} derived the abundances in an
automated fashion. They found HE~1300+0157 to have $\mbox{[Fe/H]}=-3.76$. This
called for a higher-quality spectrum for a more detailed abundance
analysis. Subsequently, the star was observed as part of the Subaru UMP
project.
 
We describe these new high-quality Subaru/HDS observations and the basic 
measurements in \S~\ref{sec:obs}. The determination of the stellar
parameters  is reported in \S~\ref{sec:stell_par}, and a description of the
abundance  analysis is presented in \S~\ref{ab_pa}. We discuss possible
scenarios for the  origin of the chemical signature of HE~1300+0157 in
\S~\ref{disc}. In  \S~\ref{sum}, we conclude with a summary of the results. 
   
\section{Observations and Measurements}\label{sec:obs} 
\subsection{Observations and Data Reduction} 
HE~1300+0157 was observed with HDS \citep{noguchi_hds} at the Subaru telescope
in two of our UMP runs. Its coordinates are $\alpha (2000) =13$\,h $02$\,m
$56.3$\,s and $\delta (2000) = +01^{\circ}\;41'\;51''$. A $0\farcs6$ slit
width was used for the red and the blue settings, yielding a resolving power
of $R\sim60,000$. CCD on-chip binning ($2\times2$) was applied. The two
wavelength settings cover the range of 3050-6800\,{\AA}. In total, 8\,h were
spent on the target, and 15\,min on the star G~64-12 (e.g.,
\citealt{1981G64-12}), as a comparison object. Individual exposures did not
exceed 45\,min in order to facilitate cosmic ray removal. See Table \ref{obs}
for more details on the observations.
 
\begin{figure} 
 \begin{center} 
\plotone{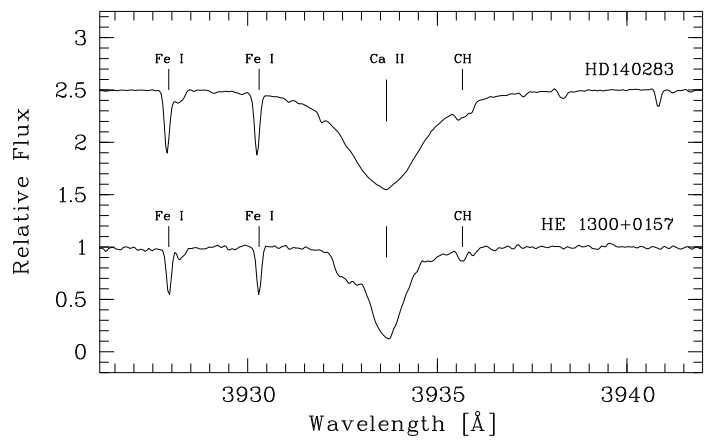}
  \caption{\label{CaK_plot} Region around the \ion{Ca}{2}\,K line in 
  HE~1300+0157 and HD~140283 ($\mbox{[Fe/H]}\sim-2.5$). Some absorption
  features  
  have been indicated. } 
 \end{center} 
\end{figure}

All of the echelle data are reduced with the IDL based software package
\texttt{REDUCE}, which is described in detail in \citet{reduce}.  Wavelength
calibration is accomplished using Th-Ar lamp frames. The reduced frames of
HE~1300+0157 are normalized using a fit to the shape of each echelle
order. After shifting the spectra to their rest frames they are weighted by
the number of counts and summed. The overlapping echelle orders are then
merged into the final spectrum. This was done for each grating setting
separately. Individual orders are kept for verification of features and
uncertainty estimates of the line measurements. 
The $S/N$ ratio of the final rebinned spectrum is $S/N\sim35$ per
$47.8$\,m{\AA} pixel at $\sim3600$\,{\AA}, $S/N\sim70$ per
$53.6$\,m{\AA} pixel at $\sim4100$\,{\AA}, and $S/N\sim170$ per
$71.6$\,m{\AA} pixel at $\sim6700$\,{\AA}.  This corresponds to a
pixel size of $\sim3.6$\,km\,s$^{-1}$.  A portion of the final
spectrum around the \ion{Ca}{2}\,K line is presented in
Figure~\ref{CaK_plot}. For comparison purposes a spectrum
\citep{norris96data} of HD~140283, a subgiant with
$\mbox{[Fe/H]}\sim-2.5$ \citep{ryan96}, is also shown.

\subsection{Broadband Photometry} 
 
In 2003, April 04 and 2006, January 17 standard $BVRI$ CCD photometry of
HE~1300+0157 was obtained with the ESO/Danish 1.5\,m telescope at La Silla
\citep{beers_photom}.  Table~\ref{pho} lists the results.  From the Schlegel,
Finkbeiner, \& Davis (1998) maps we obtain an interstellar reddening estimate
of $E(B-V)=0.022$. This is in agreement with the lower limit of
$E(B-V)=0.013$, based on the technique of \citet{munari_Na_reddening97},
described below.  To obtain reddening corrections of other passbands based on
$E(B-V)$, we made use of the relative extinctions given in
\citet{bessell_brett} and \citet{schlegel}.
 
\subsection{Radial Velocity}\label{rad} 
We used the \ion{Na}{1}\,D and the \ion{Mg}{1} b lines in the red and
the \ion{Mg}{1} triplet in the blue spectrum, as well as \ion{Fe}{1}
lines across the entire wavelength range, to measure the radial
velocity of HE~1300+0157. The heliocentric radial velocity
measurements for two epochs, averaged for every observing night, are
listed in Table~\ref{obs}. The final averaged velocity is $v_{\rm
r}=74.8$\,km\,s$^{-1}$. The standard error of this value is
$\sim0.2$\,km\,s$^{-1}$, while the dispersion is
$\sim0.4$\,km\,s$^{-1}$. There is a possible systematic uncertainty
arising from instrument instabilities of $\sim0.5$\,km\,s$^{-1}$
\citep{Aokihe1327}.  Previously, \citet{heresII} measured $v_{\rm r}=
73.6$\,km\,s$^{-1}$ from their VLT/UVES spectrum. This is in good
agreement with our value, given that they report radial velocity
uncertainties of a few km\,s$^{-1}$. From a comparison spectrum of
G~64-12, we measure $v_{\rm r}=443.1$\,km\,s$^{-1}$, which is in good
agreement with the well-established value of $v_{\rm
r}=442.5$\,km\,s$^{-1}$ by \citet{lathamG64}.

\subsection{Interstellar Absorption} 
Interstellar absorption lines are identified in both the \ion{Na}{1}\,D, and
the \ion{Ca}{2}\,H and K lines.  Since the interstellar \ion{Na}{1}\,D lines
are blended by strong telluric \ion{Na}{1}\,D emission lines that appear on
the blue side of the interstellar lines, we could only detect two
components. Component~1 is a strong, major contributer, whereas the second
component appears to be weak and located at the red side of the base of the
telluric emission lines. We measured the wavelengths of the two components and
their radial velocities relative to the laboratory scale of the star. Lower
limits are derived for the equivalent widths of the combined two components by
direct integration using the task {\tt splot} of IRAF\footnote{IRAF is
distributed by the National Optical Astronomy Observatories, which is operated
by the Association of Universities for Research in Astronomy, Inc., under
cooperative agreement with the National Science Foundation.}. As for the
\ion{Ca}{2}\,K line, we could not detect any definite components due to the
lower $S/N$ ratio (see Figure~\ref{CaK_plot}). We thus treat the feature as a
single line, and measure the same quantities as for the \ion{Na}{1}\,D
lines. The interstellar component of the \ion{Ca}{2}\,H line is too weak to be
measured accurately. The results are presented in Table~\ref{Tab:ISNa}.

The radial velocity is in good agreement with the average values of the two 
\ion{Na}{1}\,D line components. This fact clearly confirms that the absorption 
found in the blue wing of the stellar \ion{Ca}{2}\,K line is of interstellar 
origin. If we estimate the interstellar reddening $E(B-V)$ from the equivalent 
width of the \ion{Na}{1}\,D2 line, based on \citet{munari_Na_reddening97}, we 
find a lower limit of 0.013. This is consistent with the adopted value of 
0.022 derived from the \citet{schlegel} maps.

A constraint on the distance to the star can be inferred from the equivalent
widths of the \ion{Na}{1}\,D2 and \ion{Ca}{2}\,K lines. However, the
\ion{Na}{1}\,D2 line suffers from strong emission due to telluric lines.
Hence, the constraint is much weaker than can be inferred from the
\ion{Ca}{2}\,K line. According to Figure~7 of \citet{hobbs74}, the equivalent
width of the \ion{Ca}{2}\,K line provides a lower limit on the distance to
HE~1300+0157 of about $640$\,pc. This result is consistent with the 2.4\,kpc
distance, derived from the isochrone of \citet{Y2_iso} (see below).

\subsection{Line Measurements} 
For the measurements of atomic absorption lines we use a line list based on
the compilations of \citet{aoki_pasp_2002} and \citet{heresII}, as well as our
own collection retrieved from the VALD database \citep{vald}. References for
$gf$ values can be found in these papers. Equivalent width measurements are
obtained by fitting Gaussian profiles to the observed atomic lines. See Table
\ref{eqw} for the lines used and their measured equivalent widths.
 
For blended lines and molecular features, we use the spectrum synthesis
approach. The abundance of a given species is obtained by matching the
observed spectrum to a synthetic spectrum of known abundance by minimizing the
$\chi^{2}$ fit between the two spectra. The molecular line data employed for
CH are based on \citet{jorgensen_CH}, whereas the NH line data was taken from
\citet{kurucz_nh}. For OH we used the \citet{gillis_ohlinelist} line list.
Abundance uncertainties arising from this method are usually driven by
difficulties of continuum placement, and are around $0.1-0.3$\,dex.
 
\section{Stellar Parameters}\label{sec:stell_par} 

\subsection{Effective Temperature }

We have available optical $B, V, R_{\rm C}, I_{\rm C}$ (where 'C'
indicates the Cousins system) and near-infrared 2MASS $J,H,K$
\citep{2MASS} photometry. To determine the effective temperature
(\mbox{T$_{\rm eff}$}) from the available colors we employ the Alonso,
Arribas, \& Martinez-Roger (1996) temperature calibration. We use the
lowest available metallicity of $\mbox{[Fe/H]}=-3.0$, because
extrapolation may yield unphysical values \citep{ryan99}.  The Alonso et
al. calibrations require the $B-V$, $V-R$, $V-I$ and $V-K$ colors to
be in the Johnson system, while the $J-H$ and $J-K$ have to be in the
Telescopio Carlos Sanchez (TCS) system.  Where possible, we dereddened
the colors first and then transformed them into the required
system. This accounts for potential differences in the spectral energy
distributions of stars of a given reddened and dereddened color. To
transform the colors into the Johnson system, the transformations by
\citet{bessell83} (for $V-R$ and $R-I$) and \citet{bessell79} (for
$V-I$) are used. For the $V-K$, $J-H$ and $J-K$, we transformed the
2MASS $J,H,K$ magnitudes into the TCS system with the relations given
in \citet{TCS}. The $V$ magnitude is the same in any color system, and
for $K$ we assumed the same, since any difference between systems is
exceeded by the errors of the transformations. Hence, we do not
transform those magnitudes.

Table~\ref{pho} lists the individual effective temperatures obtained
from our set of colors. We discard the highest and lowest of the seven
temperatures, weight the remaining values according to their color
uncertainties, and average them. This yields \mbox{T$_{\rm
eff}=5450\pm70$}\,K for HE~1300+0157. The error of 70\,K is the
standard error of the mean value. Systematic uncertainties in the
temperature are likely to be higher, but are not further considered
here. We note here that the Alonso et al. (1996) calibrations for
$V-K$ and $J-K$ are the least [Fe/H] sensitive ones. If one were to
use just those two color indices, the average of those two
temperatures would be in the very good agreement with our value of
\mbox{T$_{\rm eff}=5450$}.

\begin{figure} 
 \begin{center} 
\plotone{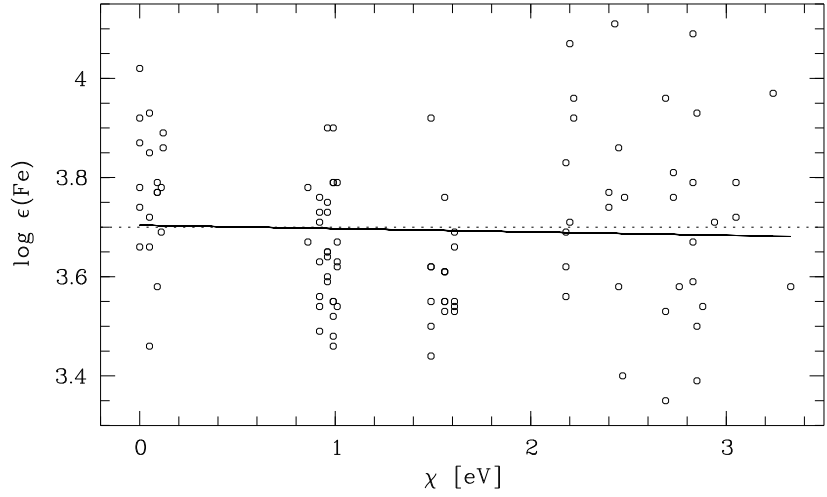}
  \caption{\label{t_excit} LTE abundances $\log \epsilon(\rm Fe)$ of all
  measured \ion{Fe}{1} lines as a function of excitation potential $\chi$ for
  the adopted subgiant stellar parameters. There is no significant trend of
  abundance found with $\chi$, indicating that the photometrically derived
  temperature is close to the value inferred from the excitation balance. }
 \end{center} 
\end{figure}

In Figure~\ref{t_excit} we test whether the photometrically derived 
temperature agrees with the temperature that can be derived from demanding no 
trend of abundances with excitation potential of the \ion{Fe}{1} lines. The 
details on the abundances will be given below. Over the range of 0 to 
$\sim3.3$\,eV our adopted \mbox{T$_{\rm eff}=5450$\,K} yields no significant 
trend of \ion{Fe}{1} abundances.

We note that \citet{heresII}, despite partially having the same set of
photometry (the 2003 data), determined a temperature slightly different (by
$\sim40\,K$) from ours (see Table~\ref{stell_par}). This is likely due to a
different procedure of transforming all the colors into the required
photometric systems.

\subsection{Microturbulence and Surface Gravity} 
 From inspection of a 12\,Gyr isochrone with $\mbox{[Fe/H]}=-3.5$ and
$\alpha$-enhancement of $\mbox{[$\alpha$/Fe]}=0.3$ \citep{green,
Y2_iso}, we find that our adopted temperature results in two
possibilities for the evolutionary status of HE~1300$+$0157 -- a
subgiant case ($\log g=3.2$) and a dwarf case ($\log g=4.6$). By using
such isochrones we are assuming that the Alonso et al. (1996)
color-temperature relations agree with those used in the construction
of the isochrones.  For each gravity possibility the microturbulence,
\mbox{v$_{\rm micr}$} was obtained from $\sim100$ \ion{Fe}{1} lines by
demanding no trend of abundances with equivalent widths. We derive
\mbox{v$_{\rm micr}\sim1.4$}\,km\,s$^{-1}$ for the subgiant case,
whereas no \mbox{v$_{\rm micr}$} could be obtained for the dwarf
case. A negative value
\footnote{We investigated this problem, and found the same effect was
occurring for HD~140283 when using a dwarf-gravity model atmosphere for this
well-known subgiant.} would be required to produce no trend of abundance with
equivalent width.  We then repeat this exercise using 40 \ion{Ti}{2} lines,
and obtain \mbox{v$_{\rm micr}\sim1.7$}\,km\,s$^{-1}$ (subgiant case) and
\mbox{v$_{\rm micr}=0.8$}\,km\,s$^{-1}$ (dwarf case). The straight average of
the Fe and Ti values is subsequently adopted for the subgiant case, and the Ti
value for the dwarf case.

We next attempt to use the ionization equilibrium of \ion{Fe}{1} and
\ion{Fe}{2} abundances to find the surface gravity. Adopting a NLTE correction
for \ion{Fe}{1} of +0.2\,dex (e.g., \citealt{asplund_araa}), we obtain a
gravity of $\log g\sim4.1$. That is precisely between the two isochrone
values. Any NLTE correction assumed for \ion{Fe}{1} always increases the
gravity compared with the LTE version. If we were to use the pure LTE
ionization equilibrium we would obtain a gravity closer to the subgiant branch
(i.e., $\log g\sim3.6$). \citet{heresII} used this same technique, and found
HE~1300+0157 to be a subgiant (see Table~\ref{stell_par}). They employed the
LTE ionization balance of both \ion{Fe}{1}-\ion{Fe}{2} and
\ion{Ti}{1}-\ion{Ti}{2}.  

The assumption of LTE has its limitations, so we decided to apply NLTE
corrections for our analysis. Taking into account the uncertainties in
\mbox{v$_{\rm micr}$} (e.g., $\sim0.3$\,km\,s$^{-1}$) when using the
ionization equilibrium method for the NLTE corrected \ion{Fe}{1} abundance,
the gravity could still be as high as $\log g\sim3.9$. This exercise, as well
as uncertainties in the NLTE corrections, shows that neither of the gravity
solutions can be excluded spectroscopically. We therefore adopt the two
gravity values derived from the isochrone, and carry out the abundance
analysis for both cases. This has the advantage that the gravity determination
is not dependent on the assumed \ion{Fe}{1} NLTE correction.

We also carry out a differential microturbulence and LTE ionization
equilibrium-gravity determination for both HE~1300+0157 and HD~140283, using
equivalent widths and $\log gf$ values for a subset of line strengths taken
from \citet{norris96data}. We find that HE~1300+0157 has a gravity close to
that of HD~140283. This may also indicate that the subgiant solution should be
favored over the dwarf case.

\subsection{Balmer Jump Analysis}
\begin{figure} 
 \begin{center} 
\plotone{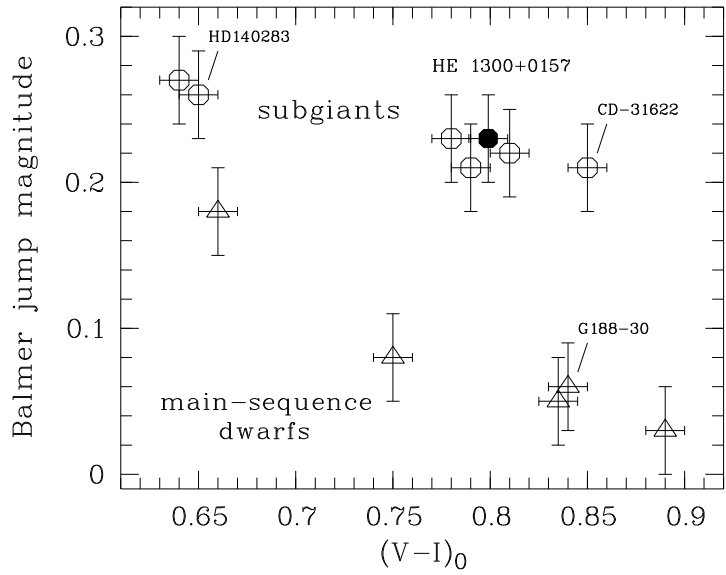}
  \caption{\label{balmer} Balmer jump magnitude as a function of 
  $(V-I)_{0}$. Open circles indicate subgiants, whereas triangles indicate 
  dwarf stars. HE~1300+0157 is shown with a filled circle.  
} 
 \end{center} 
\end{figure}

In order to make a final decision between the two gravity solutions, we
analyze the Balmer jump at 3636\,{\AA} of HE~1300+0157, and for other stars
with known surface gravities. For a given temperature and metallicity, the
Balmer jump increases with decreasing surface gravity. From flux-calibrated
medium-resolution ($\sim3-5$\,{\AA}) spectra the depth of the Balmer jump can
be measured and converted to a magnitude.  Figure~\ref{balmer} shows the
Balmer jump magnitude as a function of the $(V-I)_{0}$ color for HE~1300+0157
and the other objects. We take $b-y$ data from \citet{schuster88} and
\citet{Anthony-Twarog_by94}, and convert them to $V-I$ using a relation
obtained by Bessell \& Shobbrook (unpublished).  The predicted colors are then
dereddened.  Table~\ref{balmer_tab} lists the Balmer jump magnitude and the
dereddened $(V-I)_{0}$ color of the comparison stars. The evolutionary status
of the stars is also listed. From the relative comparison with stars such as
HD~140283 and CD~$-31^{\circ}$ 622 (subgiants) and G~188-30 (dwarf), we
conclude that the dwarf case can be excluded for HE~1300+0157. We thus adopt
the subgiant case, and HE~1300+0157 is discussed as such throughout the
remainder of the paper. This result for the surface gravity is also in
agreement with the indications (described above) from the \mbox{v$_{\rm
micr}$} determination and the comparison with HD~140283.  

Based on the 12\,Gyr, $\mbox{[Fe/H]}\sim-3.5$,
\mbox{$\mbox{[$\alpha$/Fe]}=+0.3$} isochrone \citep{Y2_iso}, we derive an
estimate of $M\sim0.82\,M_{\odot}$ for the mass of the subgiant. From
fundamental equations (Stefan-Boltzman Law, Gravitational Force) we estimate a
luminosity of $\sim 11.5$\,L$_{\odot}$ for the subgiant case. It follows that
the absolute magnitude is $M_{V}=2.1$\,mag. The resulting distance is
$d\sim2.4$\,kpc.

  
\subsection{Metallicity} 
We use 101 \ion{Fe}{1} and 7 \ion{Fe}{2} lines to derive the iron abundance
of  HE~1300+0157. Following \citet{asplund_araa}, we correct the
[\ion{Fe}{1}/H]  abundance by +0.2\,dex to account for NLTE
effects. \ion{Fe}{2} is known to be  only very weakly affected by NLTE, if
such effects are present at all (e.g.,  \citealt{asplund_araa}). The
corrected [\ion{Fe}{1}/H] abundance is then  higher than the \ion{Fe}{2}
abundance. Usually, the \ion{Fe}{2} abundance is  higher, and the NLTE
correction of \ion{Fe}{1} compensates for the  difference. Whether or not
NLTE effects for Fe are less severe in our modeling  of HE~1300+0157 than we
assume (i.e., +0.2\,dex) is not clear. However, since  the correction is of
the same order as the overall error of the \ion{Fe}{2}  abundance 
we do not regard this as a major problem.  Based on this experience, we adopt
the NLTE-insensitive \ion{Fe}{2} abundances as our final metallicity of the
star. This choice also avoids potential problems arising from the difficulties
of the \mbox{v$_{\rm micr}$} determination. The 7 \ion{Fe}{2} lines are all
weak. Hence they are expected to be not much affected by microturbulence, as
any strong \ion{Fe}{1} lines would be. In the subsequent discussion of the
metallicity of the star we thus refer to its [\ion{Fe}{2}/H] abundance. The
stellar parameters of HE~1300+0157 are summarized in
Table~\ref{stell_par}. For completeness, the [\ion{Fe}{1}/H] abundance can
also be found in Table~\ref{abund}.

\section{Abundance Analysis}\label{ab_pa} 
  For our 1D LTE abundance analysis of the Subaru spectrum we use model
atmospheres obtained from the latest version of the MARCS
code\footnote{Numerous models for different stellar parameters and
compositions are readily available at http://marcs.astro.uu.se} (B. Gustafsson
et al. 2006, in preparation). Solar abundances are taken from
\citet{solar_abund}.  

The subgiant and dwarf case abundances obtained for HE~1300+0157 are
presented in Table~\ref{abund}.  Since we believe the star to be a
subgiant, we emphasize here that the dwarf abundances should be
regarded as comparison values only. In summary, apart from a higher
metallicity, the only significant abundance differences between the
two cases are the molecular C and O abundances because they have a
higher sensitivity to the surface gravity than any values derived from
atomic lines.

\subsection{NLTE Effects} 
Despite the fact that NLTE computations may vary depending on the prescription
and/or the author (see \citealt{asplund_araa} for a discussion of this topic),
NLTE corrections should be applied to LTE abundances. This way, the best
possible abundances can be obtained, which is important for comparisons with
theoretical models. We thus provide in Table~\ref{abund} both the results from
the LTE analysis and the NLTE corrected abundances, where such corrections are
available. It is left to the reader to decide which abundances to adopt, for
example, for comparison purposes with other stars or with theoretical models,
such as those for Galactic Chemical Evolution. In any case, to avoid confusion
we will state which abundances are referred to in the following discussion.

\subsection{Lithium} 
As stars evolve from the main sequence up the red giant branch, their
surface convection zone deepens. Since the bottom of the convection
zone is hot enough to destroy the light element lithium, Li-poor
material is mixed up to the surface, diluting the surface Li
abundances. Typical main-sequence halo stars have Li abundances of
$A(\rm Li)\sim2.2$ (e.g., \citealt{spite_lithium_pl82}). Many
subgiants, however, have Li abundances lower than this value. This is
likely due to their deepening convective envelopes as the surface
temperature decreases and the stellar radius increases.

\begin{figure} 
 \begin{center} 
\plotone{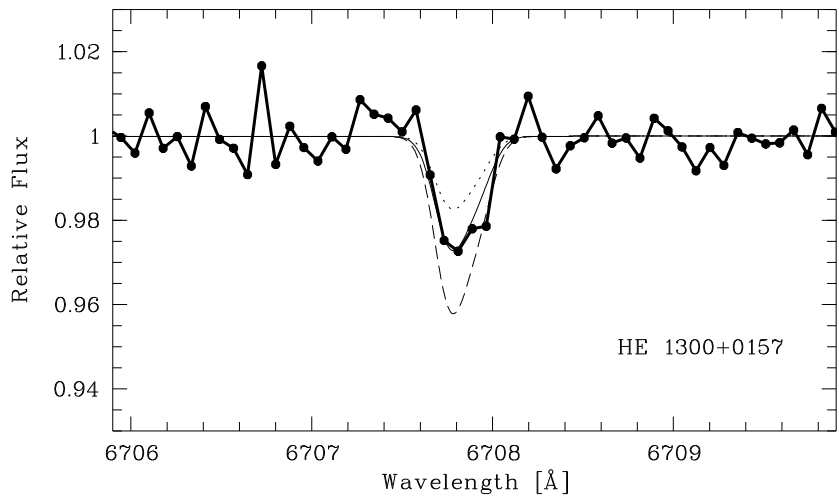}
  \caption{\label{li_plot} Spectral region of the \ion{Li}{1} doublet. The 
  observed spectrum is shown (connected \textit{dots}). Synthetic spectra with 
  abundances of $A(\rm Li)=0.9$ (\textit{dotted line}), $1.1$ (\textit{thin 
  line}) and $1.3$ (\textit{dashed line}) are overplotted.} 
 \end{center} 
\end{figure}

\begin{figure} 
 \begin{center} 
\epsscale{.6}
\plotone{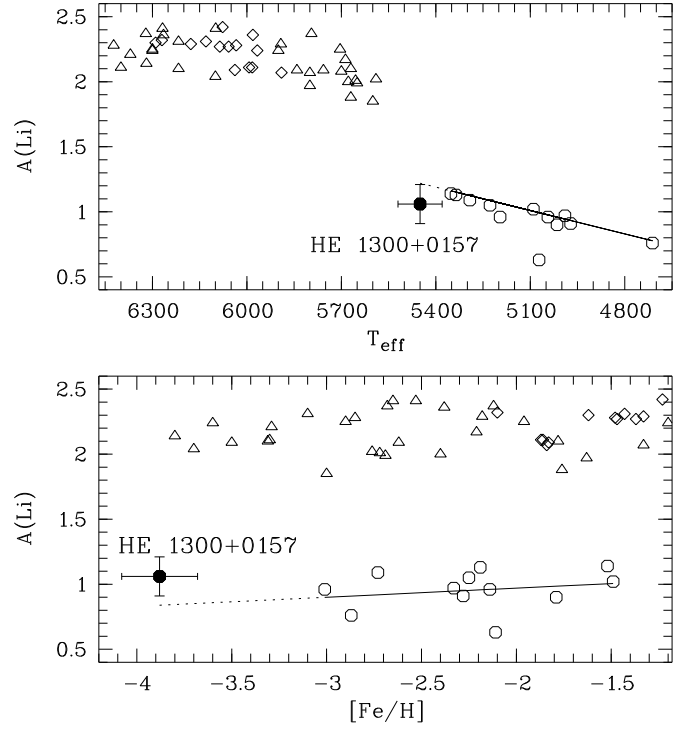}
  \caption{\label{ali_feh_plot} Lithium abundances $A(\rm Li)$ for a sample of 
  subgiants (\textit{open circles}; \citealt{garciaperez_primas2006}) and 
  selected main-sequence stars (\textit{triangles}; \citealt{RBDT96} and 
  \textit{diamonds}; \citealt{ryan_li_01}) as a function of effective 
  temperature (\textit{top panel}) and metallicity [Fe/H] (\textit{bottom 
  panel}). HE~1300+0157 is indicated with a filled circle.} 
 \end{center} 
\epsscale{1.0}
\end{figure}

We detect the \ion{Li}{1} doublet at 6707\,{\AA} in the spectrum of
HE~1300+0157.  The derived abundance is $A(\rm
Li)\sim1.1$. Figure~\ref{li_plot} shows the observed spectrum together with
synthetic spectra of different abundances. The Li abundance of HE~1300+0157 is
in agreement with values found in other subgiants (e.g.,
\citealt{garciaperez_primas2006}). This is presented in
Figure~\ref{ali_feh_plot}, where the LTE Li abundance of HE~1300+0157 is
compared with those of the metal-poor subgiant study of
\citet{garciaperez_primas2006}. Also shown are main-sequence data, taken from
\citet{RBDT96} and \citet{ryan_li_01}, to illustrate the different levels of
Li in main-sequence and subgiant stars.  HE~1300+0157 has a slightly higher
temperature (\mbox{by $\sim100$\,K}) than the hottest star of the
\citet{garciaperez_primas2006} sample. However, the warmer the star, the less
depletion in Li it should experience, so a comparison with their sample should
not cause any problems. From the top panel in Figure~\ref{li_plot} a slight
trend ($\sim0.06$\,dex \mbox{per +100\,K}) is seen, with the hotter subgiants
having higher Li abundances than the cooler ones. The Li abundance of
HE~1300+0157 roughly follows this trend.  

This leaves the question of whether the lower metallicity of HE~1300+0157 has
a more significant influence on the Li abundance. The bottom panel of
Figure~\ref{ali_feh_plot} shows the Li abundance as function of
metallicity. The subgiants from \citet{garciaperez_primas2006} have much
higher abundances compared with HE~1300+0157. However, no significant trend of
the Li abundance with metallicity is found, whether or not HE~1300+0157 is
included.  This suggests that the metallicity of the star has no significant
influence on the depletion process of Li at these low metallicities.  

\citet{garciaperez_primas2006} conclude that their subgiants have depleted Li
abundances in agreement with results from standard models of stellar evolution
(e.g., \citealt{deliyannis1990}). Given that HE~1300+0157 fits rather well
into their sample (as seen in Figure~\ref{ali_feh_plot}) we conclude that
HE~1300+0157 does not have an unusual Li abundance. This is supported by
earlier works of \citet{ryan1998}, who investigated Li depletion in stars
cooler than the stars that have Spite plateau Li abundances. Their subgiants
become significantly depleted for temperatures cooler than
$\sim5750$\,K. Their subgiants with $T_{\rm eff}\lesssim5400$\,K show the same
depletion level as the subgiants in Figure~\ref{ali_feh_plot}. On the other
hand, objects hotter than $T_{\rm eff}\sim5750$\,K have so far retained the Li
in their surface and show no signs of depletion. This effect is clearly seen
in Figure~\ref{ali_feh_plot} (top panel). From the comparison of main-sequence
and subgiant data it can thus be inferred that stars seem to experience the
most significant part of their Li depletion in a narrow temperature range of
$\sim250$\,K when their effective temperature decreases to $T_{\rm
eff}\lesssim5600$\,K.

\subsection{CNO  Elements} 
\subsubsection{Carbon} 
\begin{figure} 
 \begin{center} 
\plotone{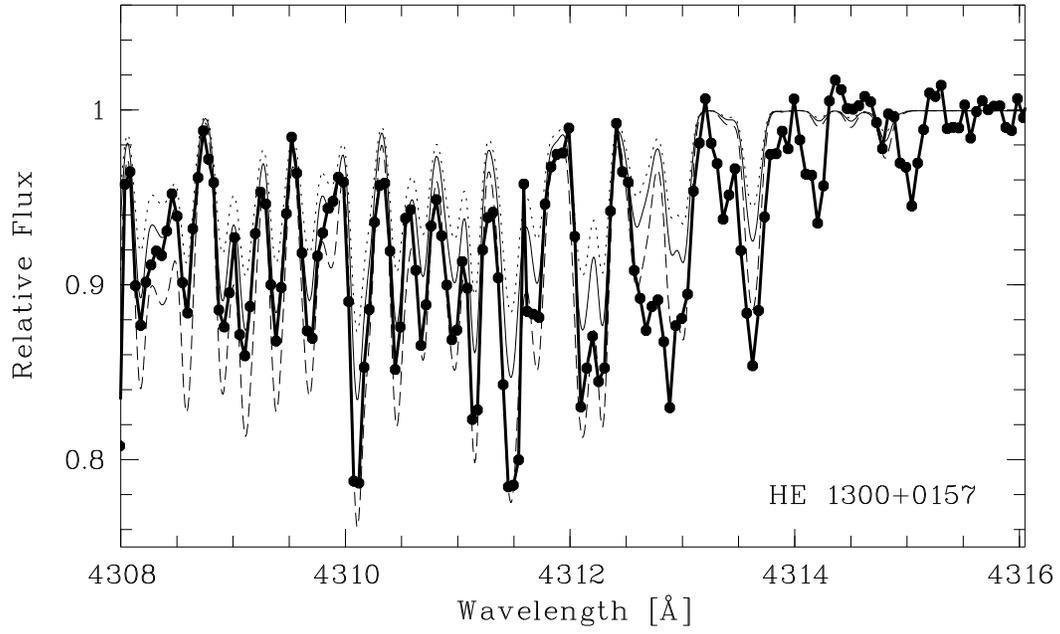}
  \caption{\label{ch4300_plot} Spectral region of the G-band band head at 
  4313\,{\AA}. The observed spectrum is shown (\textit{thin line}).  Synthetic 
  spectra with abundances of $\mbox{[C/Fe]}=1.11$ (\textit{dotted line}), 
  $1.31$ (\textit{thin line}), and $1.51$ (\textit{dashed line}) are 
  overplotted.} 
 \end{center} 
\end{figure}

\begin{figure} 
 \begin{center} 
\plotone{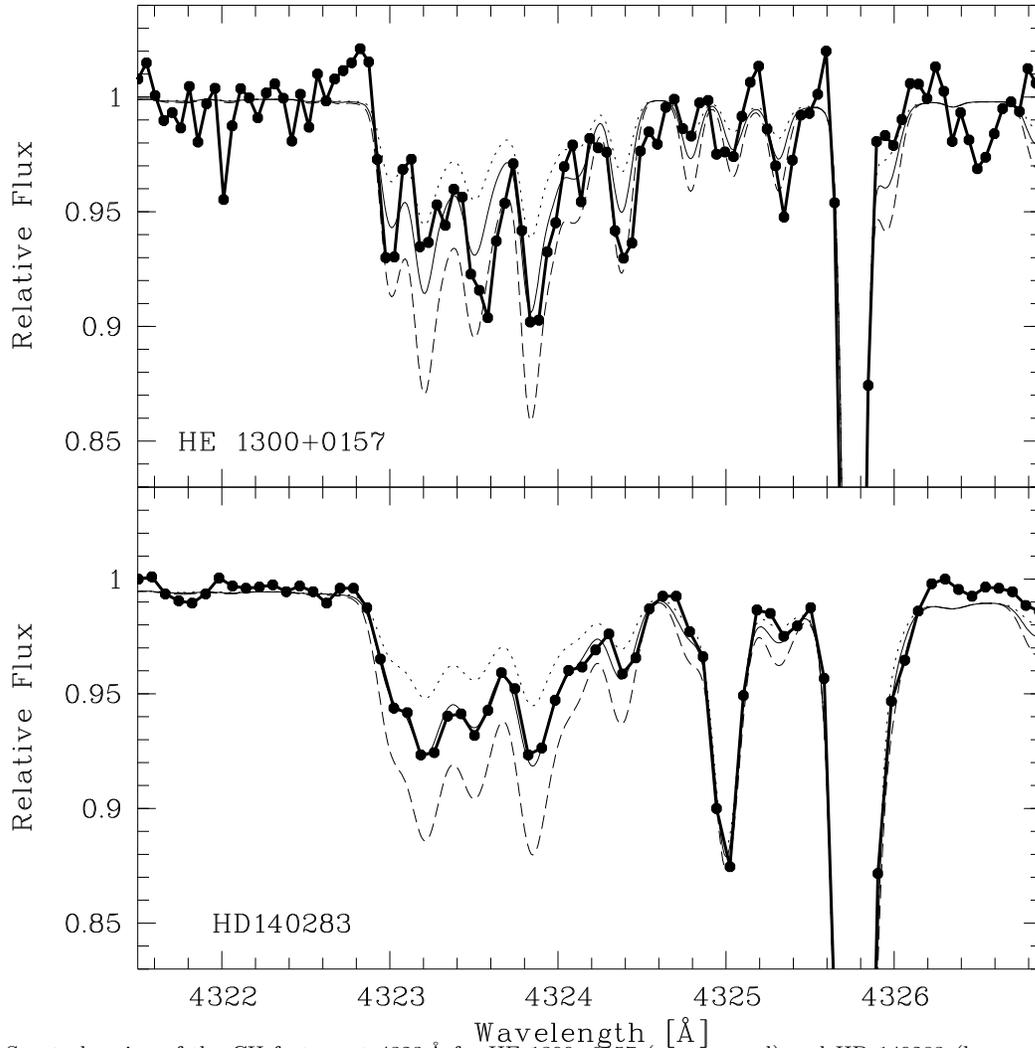}
  \caption{\label{ch4323_plot} Spectral region of the CH feature at
  4323\,{\AA} for HE~1300$+$0157 (upper panel) and HD~140283 (lower
  panel). The observed spectrum is shown (\textit{thick line}).  For
  HE~1300$+$0157, synthetic spectra with abundances of $\mbox{[C/Fe]}=1.15$
  (\textit{dotted line}), $1.35$ (\textit{thin line}), and $1.55$
  (\textit{dashed line}) are overplotted. For HD~140283, abundances of
  $\mbox{[C/Fe]}=0.34$ (\textit{dotted line}), $0.54$ (\textit{thin
  line}), and $0.74$ (\textit{dashed line}) are shown.}
 \end{center} 
\end{figure}

CH $A-X$ band features are detected between $4240$ and $4330$\,{\AA}.  Two
examples of CH features are shown in Figures~\ref{ch4300_plot} and
\ref{ch4323_plot}.  Using the spectrum synthesis approach, we measure the C
abundances from the CH feature at $4323$\,{\AA}, the G-band head
($4313$\,{\AA}), and several smaller features between $4240$ and
$4270$\,{\AA}. The main source of uncertainty for abundances derived from
molecular features is the continuum placement.  All values agree with each
other within $\sim0.25$\,dex. We adopt the straight average of the individual
measurements as our final 1D LTE C abundance.

To check the validity of our abundance measurements we also determined the C
abundance of the subgiant HD~140283. We compute synthetic spectra of different
C abundances to reproduce the CH feature at
$4323$\,{\AA}. Figure~\ref{ch4323_plot} shows this result, in comparison with
that for HE~1300+0157. We derive a 1D LTE abundance of
$\mbox{[C/Fe]}=0.54\pm0.2$ for HD~140283. Bearing in mind that we adopted
slightly different stellar parameters, this is in very good agreement with the
result of \mbox{$\mbox{[C/Fe]}=0.5\pm0.2$} derived by \citet{Norrisetal:2001}.

\begin{figure} 
 \begin{center} 
\epsscale{.55}
\plotone{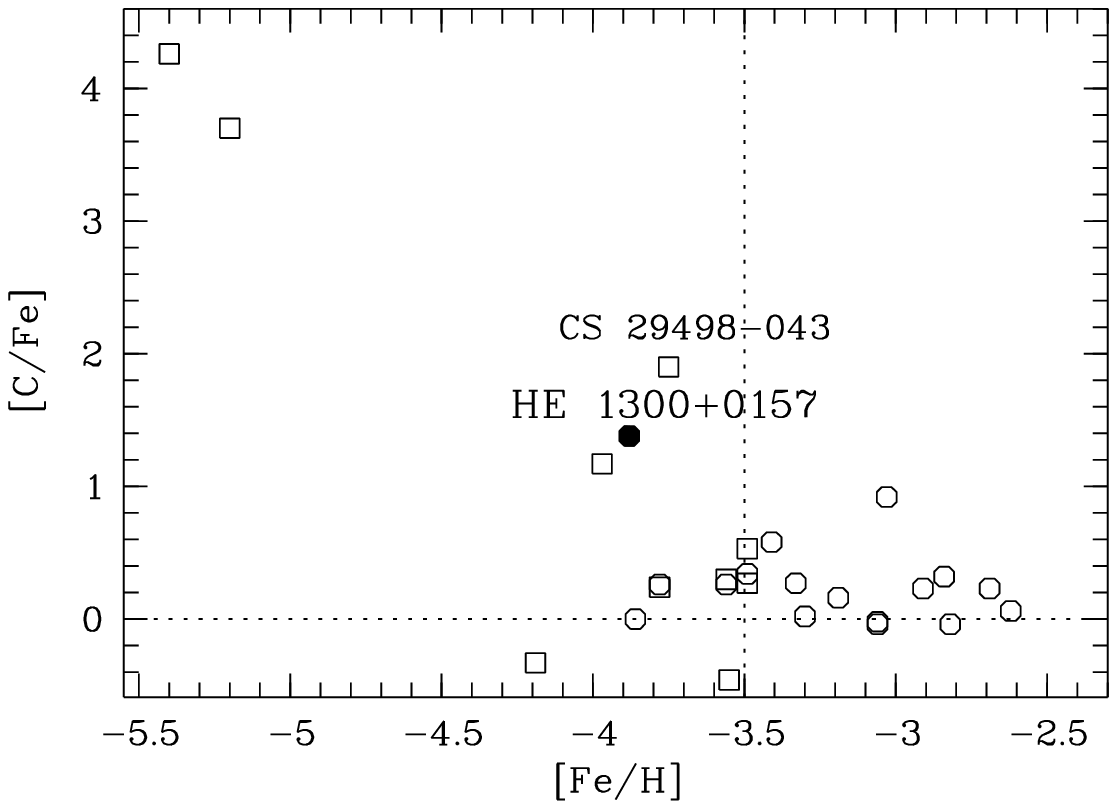}
  \caption{\label{teff_cfe} 1D LTE carbon abundance ratios as a function of
  metallicity [Fe/H] of HE~1300+0157 (\textit{filled circle}) and the unmixed
  stars of \citet{spite2005, spite2006} (\textit{open circles}).  The open
  squares refer to other stars with $\mbox{[Fe/H]}\lesssim-3.5$
  (\citealt{McWilliametal, aoki_mg}; \citealt{cayrel2004} [mixed stars];
  \citealt{cohen04, HE0107_ApJ, Aokihe1327})}
 \end{center} 
\epsscale{1.0}
\end{figure}

Figure~\ref{teff_cfe} shows the 1D LTE C abundance of HE~1300+0157, compared
with the unmixed giants of \citet{spite2006}, and additional stars with
$\mbox{[Fe/H]}\lesssim-3.5$ \citep{McWilliametal, aoki_mg, cayrel2004,
cohen04, HE0107_ApJ, Aokihe1327} as a function of metallicity [Fe/H]. We note
that since the unmixed stars are selected based on their low N abundance, the
C abundance of the those stars across the chosen metallicity range are higher
(here above $\sim$0.0), compared with a sample of mixed stars. Comparing the C
abundance of HE~1300+0157 with all of the objects having
$\mbox{[Fe/H]}\lesssim-3.5$ shows that this star is significantly more
enhanced in this element ($\mbox{[C/Fe]}\sim +1.4$) than most other
stars. However, despite the large C excess in HE~1300+0157, it is not as
extreme as in some cases, such as the most iron-deficient stars or
CS~29498-043 (e.g., \citealt{aoki_mg}). Typical values for $\mbox{[C/Fe]}$ as
derived from CH bands in objects with $\mbox{[Fe/H]}\sim-3.0$ is
$\mbox{[C/Fe]}+0.5$ (as also found in HD~140283). However, at lower
metallicity a typical value is currently not well determined. It appears
though that the C enhancement in HE~1300+0157 is at the higher end of the
[C/Fe] distribution at this metallicity. The frequently occurring large
excesses of this element at the lowest metallicities may thus reflect that the
production of C in the early Universe is due to diverse sources, and is likely
decoupled from that of other elements.

We attempted to measure the ratio of $^{12}$C/$^{13}$C. From the absence of
$^{13}$CH features in the range of $4210-4250$\,{\AA} only a conservative
lower limit of $>3$ could be inferred. Higher quality data are needed to
measure a more meaningful lower limit. Based on this result, C abundances are
computed here under the assumption that all carbon present in the star is in
the form of $^{12}$C.

\subsubsection{Nitrogen} 
We searched for the NH band at 3360\,{\AA}, but no features could be detected.
Thus, only an upper limit to the N abundance of HE~1300+0157 is derived.  

To reproduce the solar N abundance of \citet{solar_abund}, \citet{Aokihe1327}
modified the \citet{kurucz_nh} NH $gf$ values. From a comparison with the
unchanged line list, this modification resulted in an abundance correction of
$+0.4$\,dex for N. Using the uncorrected Kurucz NH line list, we are thus
required to correct our derived upper limit by $+0.4$\,dex. The final upper
limit is $\mbox{[N/Fe]}< +1.2$.  

\begin{figure} 
 \begin{center} 
\epsscale{.6}
\plotone{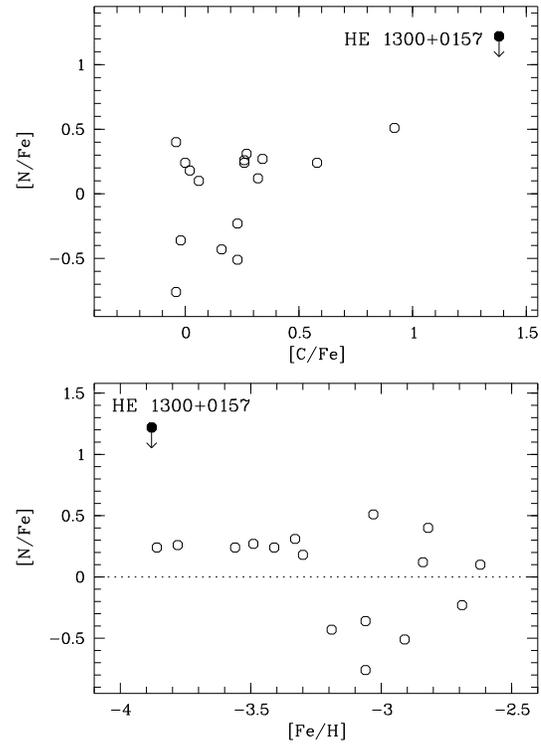}
  \caption{\label{nfe_feh} 1D LTE nitrogen abundance ratio as a function of 
  [C/Fe] (\textit{top panel}) and [Fe/H] (\textit{bottom panel}). The upper 
  limit (arrow) for HE~1300$+$0157 is indicated with a filled circle. 
The open circles refer to the unmixed giants of \citet{spite2005}.} 
 \end{center} 
\epsscale{1.0}
\end{figure}

In Figure~\ref{nfe_feh} we compare the 1D LTE upper limit of $\mbox{[N/Fe]}$
in HE~1300+0157 with the measured values of the metal-poor unmixed giants
\citep{spite2006}, as a function of [C/Fe] and [Fe/H]. The unmixed giants have
been selected by \citet{spite2006} using the requirement
$\mbox{[N/Fe]}\lesssim +0.6$.  Given that HE~1300+0157 is not a giant, it
might be expected that the N abundance of the star should not significantly
differ from the values observed in the \citet{spite2006} objects.  

Based on the C abundance and the upper limit for N, we calculate
[C+N/Fe]. Despite the availability of only an upper limit for [N/Fe], the
[C+N/Fe] ratio is very robust against changes in [N/Fe]. If [N/Fe] were solar,
the [C+N/Fe] ratio would decrease by only $\sim4\%$. Any change in this ratio
is mainly driven by the C abundance, and thus by its observational
error. Although given as an upper limit in Table~\ref{abund}, we plot this
ratio as a normal data point in our figures. Knowledge of this ``combined''
abundance makes it easier to compare HE~1300+0157 with further evolved (giant)
stars that have converted some of their C into N. We hereby assume that O is
not significantly affected by any conversion processes occurring during the
giant branch evolution.

\subsubsection{Oxygen} 
\begin{figure} 
 \begin{center} 
\plotone{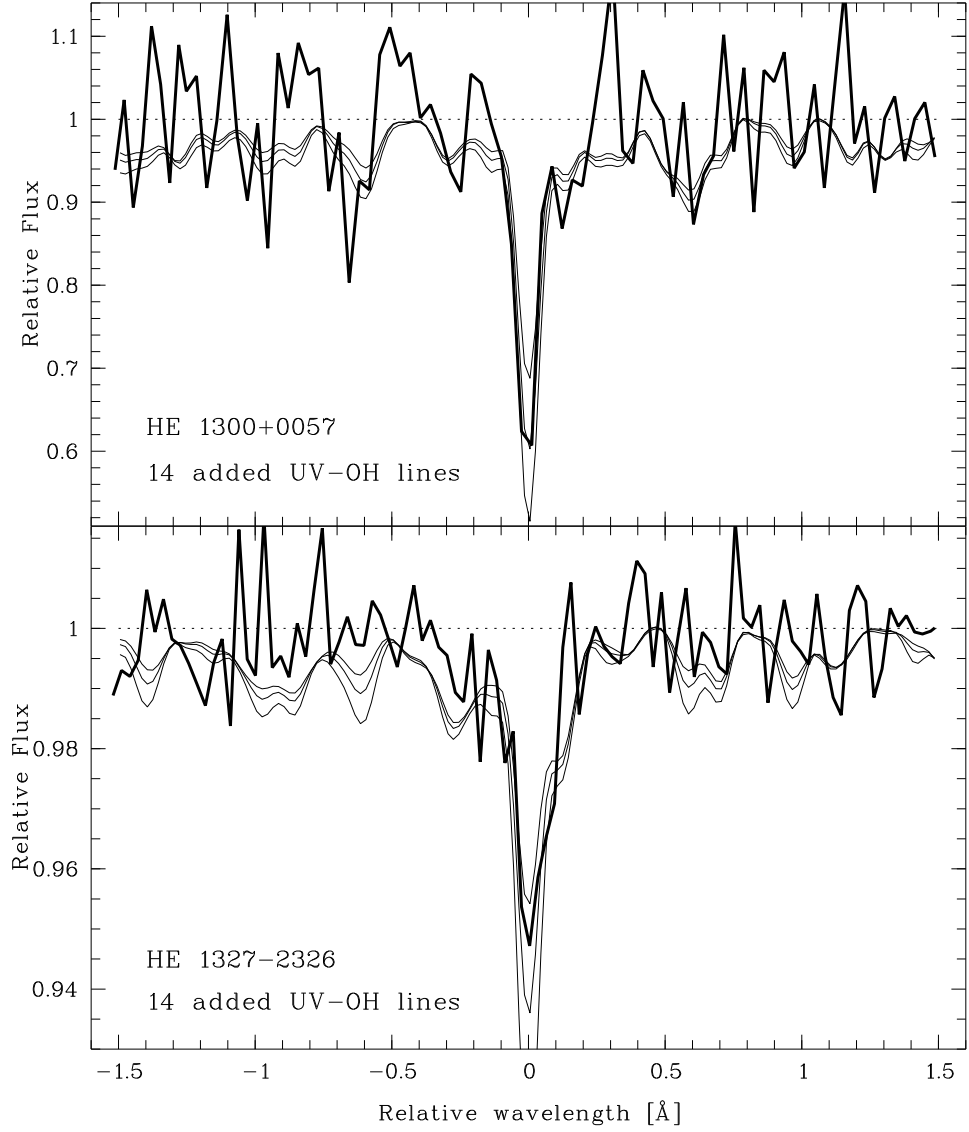}
  \caption{\label{OH_plot} Composite observed spectra (\textit{thick lines})
  consisting of 14 UV-OH lines for HE~1300$+$0157 (\textit{top panel}) and
  HE~1327$-$2326 (\textit{bottom panel}). For HE~1300$+$0157, synthetic
  spectra (\textit{thin lines}) with abundances of $\mbox{[O/Fe]}=1.56$, 1.76,
  and 1.96 are overplotted. For HE~1327$-$2326, abundances of
  $\mbox{[O/Fe]}=3.48$, 3.68 and 3.88 are shown \citep{o_he1327}.}
 \end{center} 
\end{figure}

We attempted to determine the O abundance of HE~1300+0157 from UV-OH lines
around 3130\,{\AA}. However, the $S/N$ ratio of our data in this wavelength
range is very low ($\sim5$ per pixel), and obvious identifications of OH lines
are difficult. Following \citet{norris_oxygen_method}, who co-added O
absorption features of different Ly-$\alpha$ forest redshifts in a quasar
spectrum to increase the signal, we combine 14 OH lines into a composite
spectrum. That is, we choose 14 strong OH lines, set the position of the line
in each piece to zero, and co-add the $\pm1.5$\,{\AA} wide spectra centered on
that position. We then compute a set of synthetic spectra (where OH and other
metal lines are included) and apply the same procedure. Figure~\ref{OH_plot}
(top panel) shows the composite observed spectrum together with the composite
synthetic spectra of three different O abundances ($\mbox{[O/Fe]}\sim +1.6$,
+1.8, and +2.0). The depth of the composite spectrum at 0\,{\AA} in
Figure~\ref{OH_plot} represents a $4-5\sigma$ detection, based on the $S/N$
ratio of the composite spectrum of $\sim10-15$ per pixel.  From the comparison
with the synthetic spectra, the 1D LTE abundance of HE~1300+0157 is estimated
to be $\mbox{[O/Fe]}= +1.76\pm0.3$.  

It is not clear, however, whether the observed spectrum has been normalized in
the best possible fashion. As can be seen from the synthetic spectra, OH and
other metal lines in the close neighborhood of OH lines, lower the composite
continuum significantly in a few cases. We choose to normalize the synthetic
spectrum at the composite wavelength of +0.8\,{\AA} (recalling that a
synthetic spectrum cannot exceed unity). The reason for this is due to the
(pseudo-)continuum of the composite synthetic spectrum becoming artificially
lowered with an increasing number of added synthetic spectra. We can estimate
the impact of such an effect by considering a small wavelength region that
does not have any lines -- by definition this region must be unity in the
composite spectrum. The observed spectrum is normalized by ignoring this
effect, but with the intention of obtaining, roughly, equally distributed
noise around unity. Since it is not clear what a more sensible approach with
respect to the normalization should be, we note that the derived abundance may
be slightly underestimated.  

To test the validity of the technique and the resulting O abundance for
HE~1300+0157, we apply the same procedure (using the same OH lines) to the
VLT/UVES spectrum of HE~1327$-$2326 ($\mbox{[Fe/H]}=-5.4$), for which
\citet{o_he1327} determined the O abundance directly from several detected
UV-OH lines. The result is shown in the bottom panel of
Figure~\ref{OH_plot}. The composite VLT data is overplotted with composite
synthetic spectra with the 1D LTE abundance of $\mbox{[O/Fe]}=+3.7$ taken from
\citet{o_he1327}, as well as abundances of $\pm0.2$\,dex around this
value. The composite observed data and the synthetic spectrum agree well,
demonstrating that this technique can be used to reliably determine an O
abundance. It compares favorably with the individual line measurements when
dealing with low $S/N$- ratio UV data such as we have available in the present
study. Owing to the numerous lines present in the UV that can be combined, OH
seems to be the ideal species for this method, possibly compensating for the
``hard to get'' photons in this wavelength range.  

We note, however, that the Frebel et al. (2006) 1D LTE abundance appears
slightly too high (by about $0.1-0.15$\,dex) compared with the composite
observed spectrum. This reflects the continuum normalization problem described
above. If the continuum were slightly shifted down this discrepancy would be
corrected. From this experience we conclude that the O abundance of
HE~1300+0157 is probably underestimated by $\sim0.1-0.2$\,dex. This estimate
is of the same order as the abundance uncertainty reported in
\citet{o_he1327}. We thus do not attempt to correct our derived abundance for
this effect.  

\begin{figure} 
 \begin{center} 
\plotone{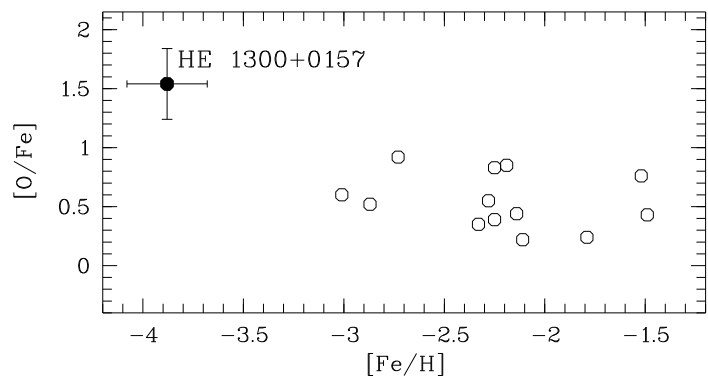}
  \caption{\label{OH_trend} Oxygen abundances [O/Fe] for a sample of subgiants 
  as a function of metallicity [Fe/H]. The open circles refer to data taken 
  from \citet{garciaperez_primas2006_O}, while HE~1300+0157 is marked with a 
  filled circle. All abundance are determined from OH features. See text for 
  discussion.} 
 \end{center} 
\end{figure}

In Figure~\ref{OH_trend} we compare HE~1300+0157 with the subgiants of
\citet{garciaperez_primas2006_O}. The O abundances of all stars are in 1D LTE
and derived from OH lines. As can be seen, HE~1300+0157 is strongly
overabundant in O compared to the other subgiants. Possible implications for
the interpretation of the abundance pattern are discussed in \S~\ref{disc}.
     
\subsubsection{3D  Effects} 
Abundances derived from molecular lines observed in metal-poor stars are known
to be significantly overestimated in 1D model atmospheres, as compared to
their 3D counterparts (see e.g., \citealt{asplund_araa} for a review on this
topic). We thus invoke a 3D correction of $-0.6$\,dex, based on
\citet{ohnlte_asplund}, for our O abundance of HE~1300+0157. Since the effects
of 3D line formation are similar for the hydrides CH, NH and OH, we use the
same correction for the C and N abundances. Due to their similar behavior,
abundance ratios amongst the three elements are reasonably robust irrespective
of whether or not a correction is employed.  However, any such 3D correction
may have uncertainties of up to $\sim0.2$\,dex (M. Asplund, private
communication). A summary of the 1D and 3D CNO abundances can be found in
Table~\ref{abund}.

\subsection{Intermediate-Mass and Iron-Peak Elements}
 
The \ion{Na}{1}\,D resonance lines at 5890\,{\AA}  are used to determine
the  Na abundance, while Al  is 
measured from the \ion{Al}{1} resonance line at 3961\,{\AA}. The other line
of  the \ion{Al}{1} doublet at 3944\,{\AA} is heavily blended by CH lines 
\citep{al_ch_blend}. Spectrum synthesis of the 3944\,{\AA} feature yields an 
abundance in good agreement with that derived from the 3961\,{\AA} line. 
Na and Al abundances, particularly when derived from the resonance
lines, are known to be very sensitive to NLTE effects. For
HE~1300+0157, we adopt the NLTE corrections for HD~140283 of
$-$0.40\,dex (for Na) and +0.50\,dex (for Al) computed by
\citet{gehren2004_nlte}. In the absence of additional computations for
stars more metal-poor than HD~140283 the adopted NLTE corrections may
be regarded as lower limits.

Several \ion{Mg}{1} lines across the spectrum are employed to derive the
Mg  abundance.  Mg is affected by NLTE, and we adopt a +0.15\,dex correction,
as  determined for HD~140283 by \citet{gehren2004_nlte}.     

The abundances derived from several lines of \ion{Ca}{1} and \ion{Ca}{2} agree 
within $\sim0.2$\,dex. Since NLTE corrections for \ion{Ca}{1} are not 
well known, we are not able to make any correction.

The Sc abundance  is derived from several \ion{Sc}{2} lines between 3350 
and 4415\,{\AA}.  
The \ion{Ti}{2} abundance is based on numerous lines in the range
$3100-4600$\,{\AA}. It is only $\sim0.1$\,dex higher than that derived from a
few \ion{Ti}{1} lines.  NLTE effects are likely to play a role in the
derivation of this abundance, but unfortunately no calculations are available
in the literature.  

The $\alpha$-elements Mg, Si, Ca, and Ti in metal-poor stars are generally
enhanced by $\sim0.4$\,dex with respect to the solar value (e.g.,
\citealt{ryan96}). HE~1300+0157 (as a subgiant) does not deviate much from
this behavior. Its [Mg/Fe] and [Ca/Fe] abundances are elevated by
$\sim0.3-0.5$\,dex, whereas [Si/Fe] and [Ti/Fe] are higher than this value.

Iron-peak elements provide strong constraints on the very first
chemical-enrichment processes, since they are products of complete and
incomplete Si-burning in SN explosions. Furthermore, their abundances relate
to the relative masses of different Si-burning regions, the mass of the
progenitor, and the explosion energy of the SN.

The Cr abundance is measured from a few \ion{Cr}{1} and two \ion{Cr}{2}
lines. The \ion{Cr}{1} abundance is $\sim0.15$\,dex lower than that of the
ionized species, with $\mbox{[Cr/Fe]}\sim-0.1$. The reason for this
discrepancy might find an explanation in NLTE effects, but in the absence of
such computations we cannot further investigate the matter.
We attempted to measure the Mn abundances from two lines, \ion{Mn}{1} at
$4030.76$ and \ion{Mn}{2} $3441.99$\,{\AA}. Hyperfine structure of Mn  is
not taken into account, due to the weakness of the lines 
($<10$\,m{\AA}). The \ion{Mn}{1} line yields a $\sim0.4$\,dex lower abundance
than the \ion{Mn}{2} line.  \citet{cayrel2004} found that abundances derived
from the \ion{Mn}{1} triplet at $\sim4033$\,{\AA} are systematically lower
($-0.4$\,dex) than those from other lines. We follow their approach, and
correct the \ion{Mn}{1} (at $4030.76$\,{\AA}) abundance by +0.4\,dex to form
our adopted Mn abundance of $\mbox{[Mn/Fe]}\sim-0.5$. This is in good
agreement with the abundance from the \ion{Mn}{2} $3441.99$\,{\AA} line.

Finally, most of the \ion{Co}{1} lines used to obtain the Co abundance are 
located between 3400 and 3500\,{\AA}. 
The same applies to the \ion{Ni}{1} lines, of which all but two have 
wavelengths between 3200 and 3500\,{\AA}. In agreement with other metal-poor 
stars, Co is enhanced by more than half a dex, whereas Ni is essentially 
solar. 
 
The final abundances are listed in Table~\ref{abund}.
 
\subsection{Upper Limits} 
Upper limits for elements for which no lines could be detected can provide
useful additional information for the interpretation of the overall abundance
pattern, and the possible origin of the star. Based on the $S/N$ ratio in the
spectral region of the line, and employing the formula given in
\citet{o_he1327}, we derive 3\,$\sigma$ upper limits for a few iron-peak (V
and Zn) and neutron-capture elements (Sr, Y, Zr, Ba, and Eu). The results are
listed in Table~\ref{abund}.  

\citet{heresII} determined the Y abundance for HE~1300+0157 based on only the
strongest Y line at 3774\,{\AA}. Unfortunately, this line falls in the gap
between the two CCDs of our blue setting. We determine a Y upper limit from
the line at 4884\, {\AA} that agrees with the abundance derived by
\citet{heresII}. However, Y lines usually are much weaker than the Sr
resonance lines in metal-poor stars. Given that the upper limit for Sr is
rather low, it may be possible that the \citet{heresII} Y abundance is
spurious, and may actually be based on a weak noise peak (P. Barklem, private
communication). We thus adopt our upper limit for the subsequent discussion.

The element Ba has been conservatively assigned an upper limit.  We
find a feature at the correct position ($4554$\,{\AA}) in the
spectrum, and from the $S/N$ ratio in the region we calculate a
$3\sigma$ detection limit that is in agreement with the abundance of
the supposed line. However, inspection of the spectrum close of the
line position suggests that this may not be a real detection.

For a NLTE correction for Sr we adopt +0.3\,dex, following
\citet{mashonkina_sr_nlte}, while we use +0.15\,dex for Ba, as derived for
HD~140283 \citep{mashonkina_ba_nlte}. The upper limits of Sr and Ba appear
rather close to the \citet{heresII} data, whereas the limit of Eu may not be
as meaningful.  

\begin{figure} 
 \begin{center} 
\plotone{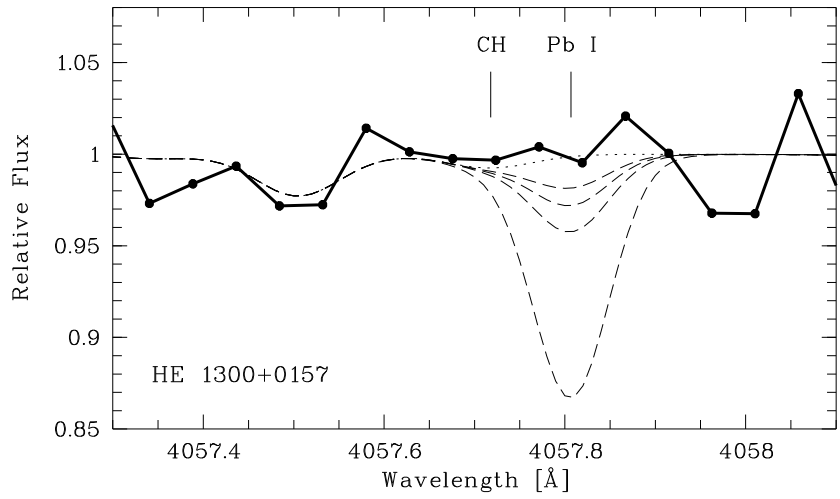}
  \caption{\label{pb} Spectral region of the \ion{Pb}{1} line. The observed 
  spectrum is shown (connected \textit{dots}). Synthetic spectra with 
  abundances of $\mbox{[Pb/H]}=-1.5,-1.3-,1.1$, and $-0.5$ are overplotted 
  (\textit{dashed lines}). The dotted line refers to a synthetic spectrum 
  where Pb is not included. Our upper limit is $\mbox{[Pb/H]}<-1.1$.  } 
 \end{center} 
\end{figure}

We searched for the strongest optical Pb line at 4057\,{\AA}, since a Pb
abundance would provide important information about the abundance pattern.
Due to potential blending with a neighboring CH line in the region of the Pb
line, it can be difficult to detect this line in very C-rich stars.  However,
in HE~1300+0157, this particular CH feature did not cause any such problems,
as can be seen in Figure~\ref{pb}, but unfortunately the line could still not
be detected. The upper limit is conservatively derived to be as low as
$\mbox{[Pb/H]}<-1.1$. This value is slightly higher than that estimated based
on the $S/N$ of the Pb spectral region.

\subsection{Abundance  Uncertainties} 
The robustness of our derived abundances is tested through changing one
stellar parameter at a time by about its uncertainty. Table~\ref{err} shows
the results for all elements.  Taking all the error sources into account, the
abundances derived from atomic lines have an overall uncertainty of
$\sim0.2$\,dex. The abundance uncertainties arising from the analysis of the
molecular features are slightly higher ($0.3-0.4$\,dex). This is mostly due to
continuum placement uncertainties, particularly in the case of O.  

Systematic uncertainties arising from the choice of color-temperature
calibration or model atmosphere are likely to add to the error
budget. Concerning the latter, our dual abundance analysis of HE~1327$-$2326
\citep{HE1327_Nature, Aokihe1327}, involving a MARCS and a Kurucz model
atmosphere, showed that the difference in abundances is less than 0.1\,dex for
most elements. We thus infer that the choice of model atmosphere is not a very
significant source of error. Uncertainties in $gf$ values have not been
considered, but are expected to be of minor influence compared with the other
uncertainties.

\subsection{Comparison  with the Barklem et al. (2005) Analysis} 
\begin{figure} 
 \begin{center} 
\plotone{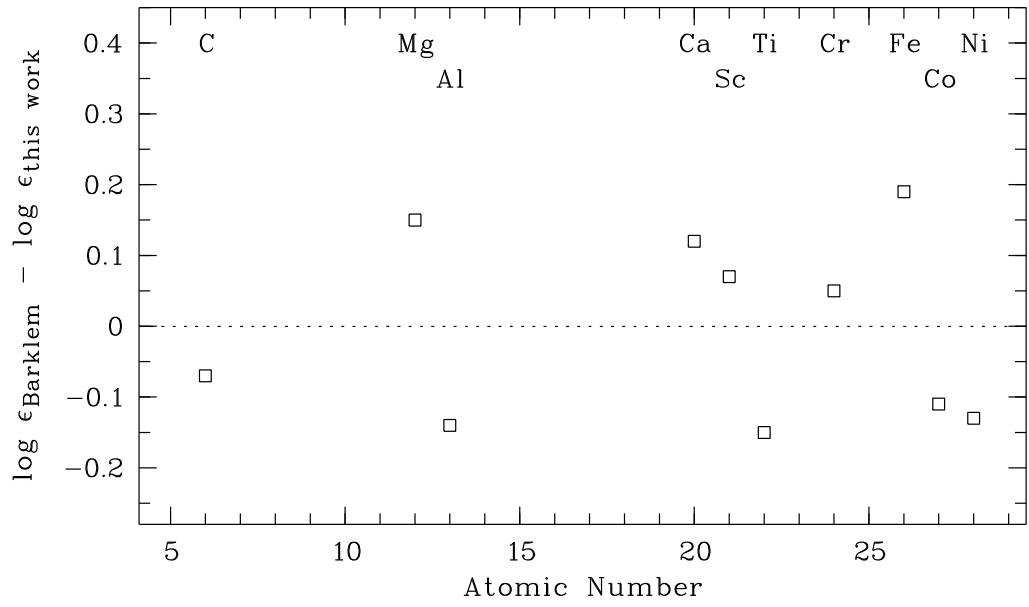}
  \caption{\label{residuals} Comparison of the abundances  of 
  \citet{heresII} with the present study. } 
 \end{center} 
\end{figure}

The abundance pattern of HE~1300+0157 was already briefly discussed by
\citet{heresII}, who analyzed the star as part of the HERES project
\citep{heresI}. These authors employed a new, automated approach to determine
homogenous sets of LTE abundances for large samples of $R\sim20,000$ snapshot
spectra of metal-poor stars. In Figure~\ref{residuals} we show the differences
of our LTE abundances compared with the LTE values obtained by Barklem et
al. (2005). The agreement is very good: the mean value is
$-0.002\pm0.04$\,dex, while the dispersion is 0.13\,dex. The differences are
likely to arise from the slightly different stellar parameters, different
abundance measurement techniques (equivalent width vs. spectrum synthesis), as
well as the use of additional lines and different atomic data.  

We emphasize here that the present study is based on significantly
higher-quality data than available to Barklem et al. (2005), and has led to
the detection of new elements in this star, including Li, O, and Na.
Furthermore, upper limits have been derived for the most important
neutron-capture elements. The new abundances (and limits) provide crucial
information for the explanation of the overall chemical abundance signature
(see \S~\ref{disc}).  As for the previously detected elements, our abundances
generally have smaller overall errors, thus refining the abundance pattern of
HE~1300+0157.
 
\subsection{Comparison with HD~140283} 
\begin{figure} 
 \begin{center} 
\plotone{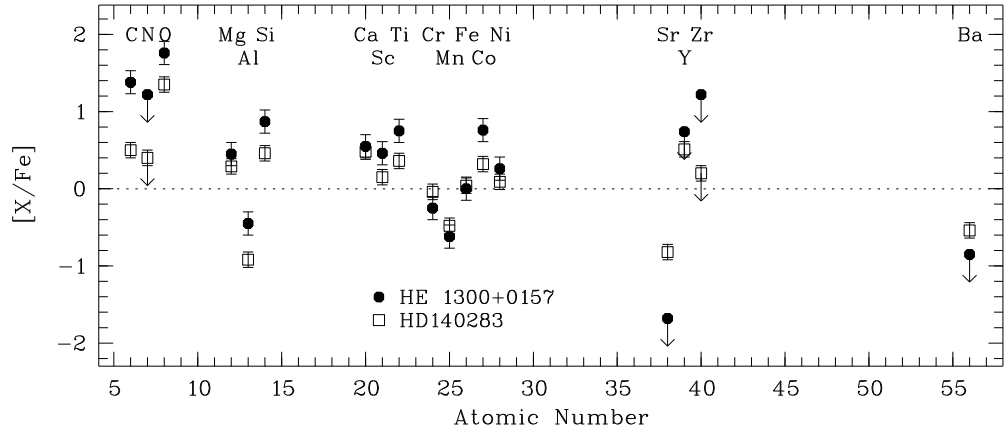}
  \caption{\label{abund_plot} LTE abundance pattern of HE~1300+0157 in 
  comparison with that of HD~140283 \citep{ryan96, boesgaard99}. The solar 
  abundances of \citet{solar_abund} have been employed for both sets of 
  abundances.} 
 \end{center} 
\end{figure}

There is a rarity of subgiants with sufficiently low metallicities to
function as direct comparison objects for HE~1300+0157. We thus
compare HE~1300+0157 with the ``classical'' halo subgiant
HD~140283. However, this object has a much higher metallicity,
$\mbox{[Fe/H]}\sim-2.5$ \citep{ryan96}. Figure~\ref{abund_plot} shows
the LTE abundances [X/Fe] with atomic number $Z$ that are measured in
the two stars. Since the abundances are scaled to Fe, the large
difference in metallicity is removed.  After the renormalization, most
of the abundances of elements with $Z<30$ appear to be very similar in
the two stars. The element obviously differing from this behavior is
C, which is significantly higher in HE~1300+0157. This indicates that
the production of C may be decoupled from that of the other
elements. 

We only have upper limits available for a few neutron-capture elements
in HE~1300+0157. They already indicate no strong
enhancements. Interestingly, the upper limits on these species in
HE~1300+0157 all seem to be at lower levels than in HD~140283. The
largest such difference is that in the Sr abundance. We note, however,
since there is a wide variety of Sr abundances observed amongst
metal-poor stars, it is difficult to speculate about the origin of
this difference in particular. The overall behaviour may suggest
though, that the elements with $Z>30$ have a different evolution from
that of the lighter elements.  This effect might possibly be related
to the lower metallicity of HE~1300+0157. The star could have been
born from gas that was not yet enriched with neutron-capture elements,
as it had occurred in the environment in which HD~140283 formed. This
is supported by the neutron-capture data of \citet{heresII}; Sr and Ba
exhibit increasing trends with increasing metallicity, [Fe/H].


\subsection{Comparison with Other Stars}
  Figure~\ref{cayrel_trend} shows the LTE abundance ratios [X/Fe] as a
function of metallicity [Fe/H] for 18 ``unmixed'' metal-poor giants
\citep{cayrel2004, spite2005}, in comparison with HE~1300+0157. We note here
that for the discussion of the abundance trends in terms of Galactic Chemical
Evolution, \citet{cayrel2004} applied NLTE corrections to Na and Al. Since we
apply slightly different NLTE corrections, we choose to compare their LTE
abundances with those of HE~1300+0157. As can be seen in the figure, the
abundances of HE~1300+0157 fit very well with the overall abundance trends of
these unmixed objects.  

\citet{cayrel2004} noted that, especially at the lowest metallicities (in the
range $-4.0\lesssim\mbox{[Fe/H]}\lesssim-3.0$), most of the abundance ratios
change only very slightly, if at all, with metallicity. Exceptions, however,
may be the cases of [Na/Fe], [Cr/Fe] and [Co/Fe]. With its low metallicity and
similar abundance pattern, HE~1300+0157 clearly supports this finding, and
provides additional constraints to the establishment of the abundance trends
at the lowest [Fe/H]. However, it is important to note that \citet{cayrel2004}
also confirm the presence of stars, such as CS~22949-037, that significantly
deviate from this well-behaved pattern. In general, below
$\mbox{[Fe/H]}\lesssim-3.5$, departures from uniformity appear much more
frequent than at higher metallicities.  

During the course of this investigation it occurred to us that the unmixed
objects are found at lower metallicities than the mixed stars in the full
sample of \citet{cayrel2004}. It is not clear whether or not this may provide
some clues to the apparently constant evolution of the elements, or if this is
simply another selection effect.  

From another comparison of similarly metal-deficient objects
\citep{Norrisetal:2001, Francois03} it appears that most of the known objects
with $\mbox{[Fe/H]}\lesssim-3.5$ have very low levels of neutron-capture
elements (e.g., Sr and Ba). The upper limits obtained for HE~1300+0157
indicate a similar behavior.  It is also worth considering the point that the
HERES survey identified no highly r-process-enhanced metal-poor stars with
[Fe/H] $< -3.2$. More stars in this metallicity range are needed to confirm
that the evolution of the neutron-capture elements was different, and possibly
repressed, in the very early Universe, as reflected by these low-metallicity
objects.

\section{Discussion}\label{disc}  
We now discuss several enrichment scenarios that may be responsible for the
overall LTE abundance pattern observed in HE~1300+0157.  We repeat that the 1D
LTE CNO abundances derived from molecular features may suffer from 3D effects
resulting in negative abundance corrections (see Table~\ref{abund}). Our
discussion is mostly based on the CNO abundances relative to each other (in
HE~1300+0157 and compared to other stars), for which 3D corrections to the 1D
values do not play such a significant role \citep{asplund_araa}. Hence, our
conclusions should not be affected.  

We note here that a ``self-enrichment'' scenario can be excluded as a
potential explanation for the origin of the abundance pattern in this
star. Due to the unevolved nature of HE~1300+0157, it is not possible that CNO
processed material was dredged up to the surface.  

Before we discuss the potential scenarios, we briefly describe how
HE~1300+0157 compares with the chemical properties of other metal-poor
objects. The grouping of similar objects can provide vital clues to the origin
of their abundance patterns.

\subsection{The CEMP-no  Group} 
The only peculiarities in the chemical signature of HE~1300+0157 are its large
C and O abundances with respect to iron and the Sun. Accordingly, the object
can be classified as a Carbon-Enhanced Metal-Poor (hereafter CEMP
[$\mbox{[C/Fe]}> +1.0$]; \citealt{ARAA}) star. There is a large variety of
CEMP stars now observed, yet details of the C production in the early Universe
that led to the formation of CEMP objects are not well understood.  

A number of CEMP stars known in the literature
\citep{norris_cempno,aoki_cempno,heresII,aoki_cemp_2006,cohen2006} have been
used to investigate the origin of this class of stars. Some of these exhibit
similar abundance signatures to those in HE~1300+0157, i.e., a very low Ba
abundance (no neutron-capture elements were detected in HE~1300+0157).  As has
been suggested before, stars with such a characteristic appear frequently
enough in the ``zoo'' of CEMP stars to warrant the definition of the class
``CEMP-no'' (referring to no strong enhancement of neutron-capture elements;
\citealt{ARAA}). CEMP-no objects have $\mbox{[C/Fe]}> +1.0$ and
$\mbox{[Ba/Fe]}<0.0$, i.e., they are not enhanced in neutron-capture elements
associated with the s- and/or r-process. A comparison with neutron-capture
data (e.g., Sr, Y) from \citet{heresII} confirms that the abundance pattern of
HE~1300+0157 cannot be associated with s- and/or r-process enhancement, but
fits the description of a CEMP-no star.  

Interestingly, Aoki et al. (2002c, 2006) and \citet{cohen2006} find
CEMP-no stars at the lower-metallicity tail
($\mbox{[Fe/H]}\lesssim-3$) of their respective samples of C-rich
objects . This may indicate that the CEMP-no phenomenon has played a
significant role in the very early Galaxy. HE~1300+0157 has the lowest
metallicity of the known CEMP-no objects, if CS~22949-037 \citep
{McWilliametal,Norrisetal:2001, Depagneetal:2002} is not counted in
the class, since it has high levels of Mg and Si), thus supporting the
view that CEMP-no stars may generally have very low metallicities.
More such objects are needed to firmly establish whether or not they
populate a distinct metallicity range.  Knowing the distribution of
the CEMP-no class should give further insight into the variety of
C-production mechanisms that enrich these stars, as well as other
groups of C-rich objects (e.g., the CEMP-r stars such as
CS~22892-052).

Recent studies (e.g., \citealt{ryan_cemps,aoki_cemp_2006,cohen2006})
investigated samples of CEMP-no stars as opposed to CEMP-s (s-process rich;
$\mbox{[Ba/Fe]}> +1.0$) stars to learn more about their origins.  From their
compilation of literature data, \citet{ryan_cemps} found that CEMP-no stars
are primarily located ``high up the first ascent giant branch'', and that they
have higher N abundance than CEMP-s objects. However, \citet{aoki_cemp_2006}
did not find such a difference in evolutionary status of the two groups on the
basis of their much larger CEMP sample.  

The abundance pattern of HE~1300+0157 agrees with the description of the
CEMP-no picture, except that it likely does not have the high N abundances
($\mbox{[N/Fe]}> +1.0$) observed in almost all of the objects found in this
class. The near-UV NH band could not be detected in HE~1300+0157, and the
upper limit is $\mbox{[N/Fe]} < +1.2$.  

We therefore suggest that this star may be an unevolved example of the CEMP-no
group which has not yet begun the conversion of its C into N. The C abundance
of $\mbox{[C/Fe]}_{\rm 1D}\sim +1.4$ should be large enough to allow for some
conversion into N during its future giant-branch evolution, producing
comparable C and N excesses to other currently observed evolved giants in the
CEMP-no class. The abundances of elements other than C, N, and O are not
expected to significantly change during the main sequence-giant branch
evolution of a star. Hence, the definition of a CEMP-no star should still be
appropriate to HE~1300+0157 at a later evolutionary stage. This may also be
the case for some of the \citet{aoki_cemp_2006} CEMP-no objects. To test this
idea, in Figure~\ref{cn_ba} we compare the [C+N/Fe] ratio of HE~1300+0157 with
the values of CEMP-s and CEMP-no stars found in the literature. We note that
only the stars with $\mbox{[C/Fe]}> +1.0$ have been chosen. Assuming that O is
not significantly processed in the way C is, the combined C+N abundance is
rather independent of the evolutionary status of the stars, and may provide
some clue to the nature of CEMP-no objects.  

In Figure~\ref{cn_ba} one sees two well-separated groups of stars, which make
up the CEMP-s and CEMP-no stars. Concerning the CEMP-no stars, some cluster
around $\mbox{[C+N/Fe]}\sim +1.0$, while the others are located around
$\mbox{[C+N/Fe]}\sim +2.2$. For the star that only has an upper limit
available for N, and thus [C+N/Fe], we calculated a lower limit by setting
[N/Fe]=0. Those values are connected in Figure~\ref{cn_ba}. We repeat here
that we plot only the upper limit for HE~1300+0157, since the lower limit is
within the observational error of the upper value. Despite the upper limit for
Ba, HE~1300+0157 fits well within the group of CEMP-no stars with
$\mbox{[C+N/Fe]}\sim +1.0$. Hence, it is likely that these stars share a
common origin.  

The other CEMP-no stars with higher [C+N/Fe] ratios, however, could have a
different origin, or they might just be extreme examples of CEMP-no stars. One
of those objects is CS~22949-037, an extremely metal-poor CEMP-no star with
some abundance signatures that are very different from the ``ordinary''
patterns observed in the stars with lower [C+N/Fe]. This object, and also
CS~29498-043, exhibits very large excesses of the $\alpha$ elements (in
particular Mg and Si; \citealt{aoki_mg}) that are not observed in other
CEMP-no stars (apart from the ``usual'' enhancements of $\sim +0.3$ to
+0.5\,dex). In fact, \citet{aoki_studiesIV} assign these two stars to a new
class of CEMP objects, the CEMP$-\alpha$ stars.  Assuming a common origin of C
in CEMP-no stars, the different levels of $\alpha$-element enhancement seen in
HE~1300+0157 and the CEMP-$\alpha$ stars may suggest a variation of the
nucleosynthesis process (e.g., caused by SNe with different mass cuts)
responsible for the abundance patterns. A similar case has been discussed by
\citet{iwamoto_science} for the two most iron-deficient stars (HE~1327$-$2326
and HE~0107$-$5240, at $\mbox{[Fe/H]}\sim-5.5$) to explain their different
levels of Na, Al, and Mg. However, since other significant differences also
exist between HE~1300+0157, on the one hand, and CS~22949-037 and
CS~29498-043, on the other, such as CNO level or neutron-capture abundances,
it is equally possible that these two types of objects indeed have a different
origin. In the absence of further CEMP-no stars with similar high Mg
abundances that can be used to test different nucleosynthesis models, no
definitive conclusion for the origin of the abundance pattern can be derived.

In summary, HE~1300+0157 is a unique CEMP star having extremely low
metallicity and no anomalous excess of alpha elements. The evolutionary status
(subgiant) is also unique, and the normal Li abundance is also
significant. Such observational facts provide new constraints on formation
scenarios of CEMP-no stars.

\subsection{Pre- Enrichment by Early SNe}
  \citet{ryan_cemps} suggested that CEMP-no objects are likely born from gas
with a large $^{12}$C/$^{13}$C ratio.  Following them, it is important to
understand how the gas of the birth clouds of CEMP-no stars could have become
enriched with large enough amounts of C. Furthermore, in HE~1300+0157, the
remaining abundances need a simultaneous explanation: the high C and O (with
lower N) abundances in combination with the ``ordinary'' appearance of
elements with $Z<30$, and possibly the underabundances of neutron-capture
elements.  

We now examine how enrichment events associated with the explosions of
different types of massive first stars could be invoked to reproduce the
abundances observed in HE~1300+0157.
 
\subsubsection{Massive Population\,III Stars} 
Non-rotating massive first stars with $M\gtrsim100M_{\odot}$ are thought to
produce copious amounts of C and O in their pre-explosion phase, but only
little N (e.g., \citealt{heger2002}). If those objects experienced at least
some mass loss before exploding as pair-instability SNe, they could pollute
the interstellar medium with corresponding amounts of these elements.  Models
of $60\,M_{\odot}$ stars that, on the contrary, explode as core-collapse SNe
\citep{meynet2005}, also produce much more C and O than N, and have some mass
loss during their evolution.

Since it is not clear how low the N abundance is in HE~1300+0157 (only an
upper limit is available), the possibility of such massive stars being
responsible for the CNO pattern could depend on the stellar rotation of the
progenitor (e.g., \citealt{fryer2001, meynet2002}). Non-rotating models
generally produce only little N (e.g., \citealt{woosley_weaver_1995,
UmedaNomoto:2002}), while rotation significantly increase the N production as
well as the mass loss (e.g., \citealt{meynet2005}). Hence, if there was a
range of rotational velocities present among such massive $Z\sim0$ stars, it
appears possible that slowly rotating massive objects might have ejected CNO
elements with similar ratios to those observed in HE~1300+0157.
 
\subsubsection{Different  Types of SNe} 
\citet{UmedaNomoto:2002} explored the chemical yields of a range of
progenitor  masses ($M\sim10-40\,M_{\odot}$) to compare them with the
observed abundance  patterns of metal-poor stars.  They have shown that a
variety of chemical  patterns of stars with $\mbox{[Fe/H]}\lesssim-3.0$ can
be modeled from the  ejecta of different Population\,III SNe (see also
\citealt{UmedaNomoto:2005},  and \citealt{nomoto2006} for a recent review on
this topic):   
\begin{itemize} 
\item[1a.] Metal-poor stars with $\mbox{[Fe/H]}<-3.5$ that have no 
overabundances of CNO elements are suspected to have formed from gas enriched 
by so-called hypernovae (with $E\gtrsim10{\times}10^{51}$\,ergs). 
 
\item[1b.] Abundance patterns of metal-poor stars with large excesses of CNO 
elements can be explained in terms of ``faint'' SNe experiencing mixing and 
large fall-back. 
 
\item[2.] Normal SNe (characterized by explosion energies of 
$E\sim1{\times}10^{51}$\,ergs) are thought to produce abundance patterns 
observed in stars with $-3.5\lesssim\mbox{[Fe/H]}\lesssim-3.0$. 
 
\item[3.] Stars more metal-rich than $-3.0\lesssim\mbox{[Fe/H]}\lesssim-2.5$
are believed to have formed from well-mixed gas enriched by several SNe.
 
\end{itemize}

\subsubsection{Hypernovae vs.  Faint SNe} 
  As listed above, in order to reproduce the abundance pattern of an ordinary
metal-poor star with $\mbox{[Fe/H]}\lesssim-3.5$ with no chemical
peculiarities (i.e., no enhancement of CNO elements), \citet{nomoto2006}
invoked a hypernova model. 

\begin{figure} 
 \begin{center} 
\plotone{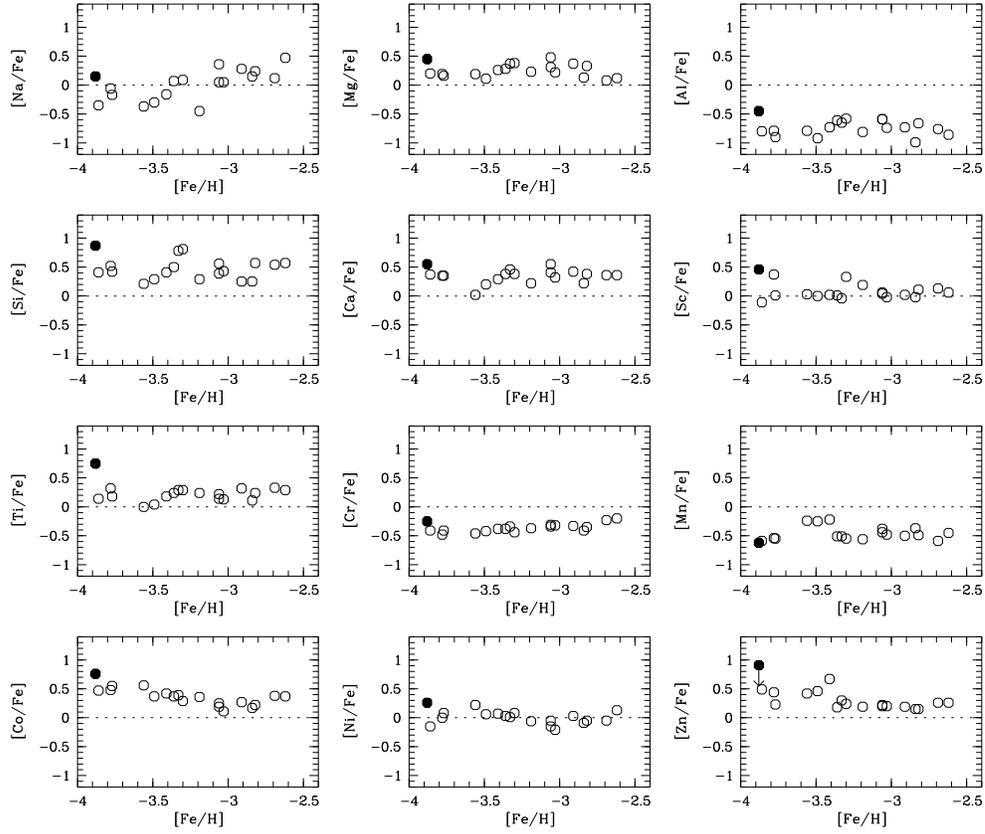}
  \caption{\label{cayrel_trend} 1D LTE abundances $\log \epsilon$ versus
  [Fe/H]  
  for elements whose abundances were measured in HE~1300+0157 (\textit{filled 
  circle}) and 18 ``unmixed'' stars (\textit{open circles}; 
  \citealt{cayrel2004, spite2006}). See text for discussion.} 
 \end{center} 
\end{figure}

\begin{figure} 
 \begin{center} 
\plotone{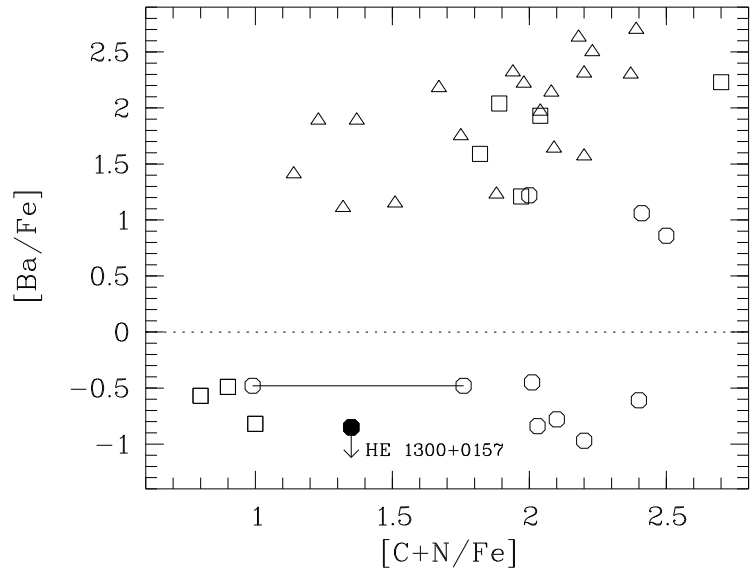}
  \caption{\label{cn_ba} Ba abundance ratio as a function of [C+N/Fe].  Open 
 symbols refer to stars with $\mbox{[C/Fe]}>1.0$ and are taken from 
 \citet{norris_cempno,Norrisetal:2001}, \citet{aoki_mg, aoki_cempno, 
 aoki_cemp_2006} and \citet{cohen2006}. Circles indicate stars with 
 $\mbox{[Fe/H]}\le-3.0$, squares indicate $-3.0<\mbox{[Fe/H]}\le-2.7$ and 
 triangles denote $\mbox{[Fe/H]}>-2.7$. HE~1300+0157 is indicated with a 
 filled circle. See text for discussion.} 
 \end{center} 
\end{figure}

With the yields of a 20\,M$_\odot$ progenitor object exploding with an
energy of $E=10{\times}10^{51}$\,ergs, they are able to fit the
averaged abundances of four typical metal-poor giants with
$-4.2<\mbox{[Fe/H]}<-3.5$ \citep{cayrel2004}. Given that the
abundances of elements in HE~1300+0157 beyond the CNO elements agree
very well with those four stars (see Figure~\ref{cayrel_trend}), it
seems that HE~1300+0157 may have been enriched by a hypernova. A
crucial test for this scenario would be an accurate measurement of the
Zn abundance. So far, we are only able to derive a $3\sigma$ upper
limit ($\mbox{[Zn/Fe]}< +0.9$). Zn strongly constrains the depth of
the mass cut and the explosion energy of the
SN. \citet{UmedaNomoto:2002} found that the large [Zn/Fe] ratios
typically observed in metal-poor stars
($0.4\lesssim\mbox{[Zn/Fe]}\lesssim0.7$; e.g., \citealt{cayrel2004})
can only be reproduced by powerful hypernovae.  The abundances of Co
and Zn are correlated because both elements are synthesized in the
same process \citep{nomoto2006}.  As can be seen in
Figure~\ref{cayrel_trend}, Co in HE~1300+0157 agrees very well with
the Co data of the \citet{cayrel2004} stars. Based on this, it may be
possible that HE~1300+0157 has a Zn abundance within the observed
range of the other stars in the same metallicity range. However,
further spectroscopic observations are needed to determine the Zn
abundance.

The only difference between HE~1300+0157 and the \citet{cayrel2004} objects
are its excesses of C and O (we assume that the N abundances of all stars are
at a similar level). This indicates pre-enrichment by a faint SN with a mixing
and fallback mechanism. Depending on the pre-SN evolution of the progenitor,
the production of N to be later ejected by the SN may greatly vary
\citep{iwamoto_science}. If such N production is not very efficient, it is
possible that a low N abundance is produced, close to the one observed in
HE~1300+015.

\subsubsection{A Possible Explanation?} 
It may have been possible that hypernova progenitors formed at the same time
as more massive Population\,III stars. In such a gas cloud a hypernova may
thus have exploded, leaving behind the chemical signature (for elements having
$Z<30$) from which ordinary metal-poor stars with $\mbox{[Fe/H]}<-3.5$
formed. If a slow-rotating very massive star would have previously contributed
to this signature with high levels of C and O elements, it may be possible to
interpret the abundance pattern of HE~1300+0157 as a superposition of these
two enrichment events. This idea would also support the formation of low-mass
stars in the early Universe from interstellar gas mainly cooled by C and O
supplied by such massive first stars \citep{brommnature}.


\subsection{Binary Mass-Transfer Scenario} 
To explain the origin of the large C overabundance, a binary scenario
might also be invoked. For CEMP-s objects, it has been shown that mass
transfer from a former AGB companion is responsible for the observed
pattern of neutron-capture elements associated with the s-process in
combination with large amounts of C. The great majority of these stars
exhibit radial-velocity variations possibly reflecting binary
membership \citep{lucatello2005}. Application of the same scenario to
CEMP-no stars does, however, cause problems. By definition, these star
have low levels of neutron-capture elements that are not predicted by
AGB-nucleosynthesis calculations (e.g., \citealt{gallino1998}).

In more general terms of explaining CEMP-no objects, however, there may be
other variations of a binary scenario in which only C, and no other elements,
are transferred.  \citet{cohen2006} have recently presented a detailed
discussion of possible binary scenarios for stars with abundance patterns
similar to HE~1300+0157. We briefly repeat the main ideas involving special
cases of AGB nucleosynthesis, and refer the reader to the Cohen et
al. discussion for further details. Where possible, we also tested whether
HE~1300+0157 could be explained in terms of these ideas.  

One possibility is that, at very low metallicity, the s-process occurring in
the former primary may run to completion to produce large amounts of Pb
\citep{busso_gallino_AGB1999}, due to a large number of neutrons per
iron-seed. As a consequence, only a small amount of Ba would remain.
\citet{cohen2006} predicted that a low-metallicity companion of such a star
should indeed display a large Pb abundance ($\mbox{[Pb/H]}=-0.5$ for a star
with $\mbox{[Fe/H]}=-3.5$).  The upper limit for HE~1300+0157 is
$\mbox{[Pb/H]}<-1.1$ (see Figure~\ref{pb}), well below what Cohen et
al. predict.
Based on this comparison, we suggest this scenario can be excluded as the
explanation for the origin of the abundance pattern in HE~1300+0157.  Another
case may be that the neutron flux is very low, thus preventing the full
operation of the s-process, and reducing the amount of Ba produced. A
low-metallicity environment may play a vital role in the creation of the low
neutron flux. However, exact details of this process are unclear at present.

Yet another binary possibility might be that the observed star is the
companion of an extremely metal-poor ($\mbox{[Fe/H]}<-2.5$) AGB star
with a mass of $M>3.5M_{\odot}$ \citep{komiya}. Such objects have a
low efficiency of the radiative $^{13}{\rm C}$ burning that does not
facilitate the production of s-process material. Mass transfer in such
a system would thus be limited to mainly C and N and would leave the
remaining abundances of the observed star unchanged.  Depending on the
mass of the AGB star large amounts of C could be converted into N (up
to the equilibrium value of the CN cycle) via hot bottom burning
occurring in the envelope. The apparently low N abundance of
HE~1300+0157 may possibly provide some constraint on the mass of the
comparison. The influence of the low metallicity on these peculiar
cases of AGB nucleosynthesis may, however, offer an explanation as to
why the CEMP-no stars have lower metallicities than the CEMP-s
objects.

%


Unfortunately, no general constraint on the binary scenario can be
derived from the detection of Li in HE~1300+0157 at a level in
agreement with its subgiant status. However, a significant amount of
mass transfer across a binary system can be excluded. Such an event
would have a negative impact on the surface Li abundance, which is
clearly not observed. However, the carbon excess in HE1300+0157 could
also be explained by the accreation of small amounts of very C-rich
material; for instance, if the accreted material has
$\mbox{[C/H]}=-0.5$, accretion of $1\%$ mass of the surface convective
layer results in $\mbox{[C/H]}=-2.5$. In such a case the surface Li
abundance is not significantly modified by the accretion. It is not
clear how the non-detection of the neutron-capture elements in
HE~1300+0157 would be explained in this special case, nor how the
large O abundance could be accounted for.

Finally, the three radial velocity measurements taken by the present
study and \citet{heresII} agree very well with one another, indicating
that HE~1300+0157 may not be a member of a binary system. The
velocities differ by only 1.2\,km\,s$^{-1}$ (see \S~\ref{rad}), which
is not a significant difference, given measurements errors. The three
measurements are 11 and 9 month apart, hence a few well-timed radial
velocity measurements of the star should easily clarify whether or not
it is a binary.


\section{Summary and Concluding Remarks}\label{sum} 
We have carried out a detailed abundance analysis of HE~1300+0157, a
subgiant with \mbox{$\mbox{[Fe/H]}=-3.9$}. The star was selected from
the HERES project \citep{heresI}, and a ``snapshot'' spectrum was
previously analyzed by \citet{heresII} in a fully automated
fashion. For the present investigation, we obtained a new high-quality
Subaru/HDS spectrum ($R\sim60,000$, $S/N\sim70$ at 4100\,{\AA}).
Using a one-dimensional LTE model atmosphere, we determined abundances
of 15 elements and upper limits for a few more. Where available, 3D
and/or NLTE corrections have been applied. Both 1D LTE and the
corrected values are presented.  Our 1D LTE abundance results agree
very well with those of \citet{heresII}.

For the measurement of the O abundance from UV-OH lines in HE~1300$+$0157, we
employed a technique that has previously been used in conjunction with
different Ly-$\alpha$ forest redshifts in quasar spectra
\citep{norris_oxygen_method}. That is, several OH lines were combined into a
composite spectrum to increase the signal strength. This was necessary because
the $S/N$ of our data was not sufficient for an individual OH line detection
around 3130\,{\AA}. To validate this technique, we applied it to
HE~1327$-$2326 to reproduce its O abundance. The result is in very good
agreement with the individual OH line measurement of \citet{o_he1327}. It was
thus shown that this technique is a powerful tool for confidently determining
an O abundance by making the best possible use of UV data with insufficient
$S/N$ ratio.  

Overall, we find HE~1300$+$0157 to be enriched in C ($\mbox{[C/Fe]}_{\rm
1D}\sim +1.4$) and O ($\mbox{[O/Fe]}_{\rm 1D}\sim +1.8$). Nitrogen could not
be detected, but the upper limit ($\mbox{[N/Fe]}_{\rm 1D}< +1.2$) shows that N
is less enhanced than C and O. All other elements are at ``normal'', low,
levels, in agreement with many ``ordinary'' very metal-poor halo stars (e.g.,
\citealt{cayrel2004}). No neutron-capture elements could be detected, and the
derived upper limits on these species indicate no significant
enhancements.

We detect Li in this subgiant star; the value is lower than the Spite-plateau
value, as expected. A comparison with other subgiants
\citep{garciaperez_primas2006} and main-sequence dwarfs indicates that stars
experience the most significant part of their Li depletion in a very narrow
temperature range around $\sim5600$\,K. On the other hand, we find that
metallicity does not seem to have a strong influence on the Li abundance in
the subgiants.  

Only a handful of objects are known at such low metallicity. Thus,
HE~1300+0157 adds important information to the puzzle of the formation
of first generations of stars in the Universe. In particular, this
star provides constraints on the production and evolution of C in the
early Galaxy, because of its large C excess and low metallicity. A few
stars have recently been found (e.g., \citealt{aoki_cempno,
cohen2006}) with a similar pattern, e.g., high C and low
neutron-capture-element abundances (the ``CEMP-no'' group). We suggest
that HE~1300$+$0157 is a less evolved member of this group. What
appears from these few stars is that they generally have a lower
metallicity compared to the objects with high(er) levels of
neutron-capture elements. This finding may provide some answers to the
production of C in the early Galaxy, but more extremely
low-metallicity stars such as HE~1300+0157 are needed to confirm this
apparent trend.

Concerning the origin of the CEMP-no group, HE~1300+0157 could indicate that
CEMP-no stars occur as single objects, rather than in binary systems. So far,
radial-velocity measurements are available at only two epochs, but they do not
indicate any variations. This has implications for what is regarded to be the
main driver of the observed abundance patterns. However, for stars other than
HE~1300+0157, there also seem to be indications for a binary model
\citep{cohen2006} to produce a similar abundance signature. Whether or not
these objects are linked in some way is not clear.  

Several potential chemical enrichment scenarios that might account for
the observed abundance signature in this star were discussed in
detail. It appears most likely that the high levels of C and O were
produced prior to the birth of the star. Our observed object would
then be a second- or later-generation star that inherited the chemical
signature of a previous-generation SNe. It is not clear exactly which
type of SN was responsible for such pre-enrichment. We speculated that
both a slow-rotating very-massive star (e.g., \citealt{heger2002}) and
a hypernova (e.g., \citealt{nomoto2006}) may have enriched the gas
from which HE~1300+0157 formed. This way, the low N ([N/Fe]) abundance
could possibly be explained together with the high C and O abundances
(arising from the slow-rotating massive star) and the normal levels of
the remaining elements (arising from a hypernova). However, a faint SN
\citep{UmedaNomoto:2002} with an unusual mass cut allowing for
excesses in C and O may also be responsible for the observed abundance
pattern in HE~1300+0157. The normal Li abundance observed in the star
supports the pre-enrichment interpretation.

Unfortunately, no pre-enrichment variation provides an immediate explanation
for all the abundance ratios we have reported. It thus remains open whether
just one such model would be able to explain an unevolved CEMP-no star such as
HE~1300+0157. {It appears that there must be several pathways for the C
production in the early Universe.} The analysis of additional CEMP-no stars
will provide important constraints on these different mechanisms.

Other possibilities for the origin of the abundance pattern might rely on the
star being a member of a binary system. However, the ``usual'' binary system
scenario includes mass transfer of C, accompanied by high levels of s-process
abundances. This is not observed in HE~1300$+$0157.  As has been pointed out
by \citet{cohen2006}, alternative explanations invoking possibilities of
special s-process-poor, C-rich mass transfer between the companions could
generally be responsible for CEMP-no objects. It is possible that if only very
little of such material was transferred onto HE~1300+0157, the surface Li could
be maintained at the observed level. However, one of these scenarios would
result in strong enrichment of this star in Pb, which could be excluded by
means of our measured upper limit. Also, a binary model could be strongly
supported by the observation of radial-velocity variations, but no such
variation has yet been found. Long-term radial velocity monitoring is thus
required to further constrain this class of scenarios as explanations for the
observed abundance pattern.

\acknowledgments A.F. thanks N. Piskunov for help with the data reduction and
acknowledges hospitality by the Uppsala Astronomical Observatory where the
reduction was carried out. Model atmosphere calculations by K.~Eriksson are
greatly appreciated. M.~Fujimoto is thanked for useful discussions about CEMP
stars. This research made extensive use of the Vienna Atomic Line Database
(VALD). A.~F., M.~S.~B. and J.~E.~N. acknowledge support from the Australian
Research Council under grant DP0342613. T.~C.~B. is supported by the US
National Science Foundation under grant AST 04-06784, as well as from grant
PHY 02-16783; Physics Frontier Center/Joint Institute for Nuclear Astrophysics
(JINA). N.~C. acknowledges financial support by Deutsche
Forschungsgemeinschaft through grants Ch~214/3 and Re~353/44. He is a Research
Fellow of the Royal Swedish Academy of Sciences supported by a grant from the
Knut and Alice Wallenberg Foundation.

{\it Facilities:} \facility{Subaru(HDS)}      


\clearpage 

\begin{deluxetable}{rrrrr} 
\tablecolumns{5} 
\tablewidth{0pt}
\tablecaption{\label{obs} Subaru/HDS observations of HE~1300$+$0157} 
\tablehead{ 
\colhead{Date}&\colhead{UT\tablenotemark{a}}& \colhead{Setting} & 
\colhead{$t_{\rm exp}$} & \colhead{$v_{\rm rad}$}\\ 
\colhead{}    & \colhead{}& \colhead{}    & \colhead{(min)} 
& \colhead{(km\,s$^{-1}$)} } 
\startdata 
2004 05 30 & 07:37 & 4030--6800\,{\AA} & 30 & 75.3 \\ 
2004 06 01 & 05:39 & 4030--6800\,{\AA} & 270& 74.6 \\ 
2005 03 02 & 12:08 & 3000--4600\,{\AA} & 180 & 74.6  
\enddata 
\tablenotetext{a}{At beginning of observation.} 
\end{deluxetable} 
 
\begin{deluxetable}{lrrrr} 
\tablecolumns{4} 
\tablewidth{0pc} 
\tablecaption{\label{pho} Photometry and the derived effective temperatures} 
\tablehead{\multicolumn{2}{l}{Magnitude/Color} & \colhead{$\sigma$}&  
\colhead{$T_{\rm eff}$}& \colhead{$\sigma_{T_{\rm eff}}$}} 
\startdata 
$V$  &14.060&0.004&\nodata&\nodata\\ 
$B-V$& 0.476&0.009&   5914&40\\ 
$V-R$& 0.382&0.007&   5611&60\tablenotemark{a}\\ 
$V-I$& 0.799&0.006&   5403&30\tablenotemark{a}\\ 
$R-I$& 0.417&0.009&   5243&60\tablenotemark{a}\\ 
$V-K$& 1.709&0.029&   5576&50\\ 
$J-H$& 0.390&0.036&   5216&140\tablenotemark{a}\\ 
$J-K$& 0.458&0.040&   5386&160\tablenotemark{a} 
\enddata 
 
\tablecomments{$BVRI$ colors are given in the Cousins photometric 
  system. $JHK$ data obtained from 2MASS. The \citet{alonso_ms} calibration 
  was used for $\mbox{[Fe/H]}=-3.0$. See text for discussion. } 
  \tablenotetext{a}{A slightly larger photometry error was used to take color 
  transformation uncertainties of $\sim0.005$\,mag into account} 
\end{deluxetable}

\begin{deluxetable}{lllr}  
\tablecolumns{7}  
\tablewidth{0pt}  
\tablecaption{\label{Tab:ISNa} Interstellar \ion{Ca}{2}\,K and  
\ion{Na}{1}\,D Lines in the spectrum of HE~1300+0157}  
\tablehead{  
\colhead{Component} &  
\colhead{$\lambda$\tablenotemark{a}}& 
\colhead{$v_{\rm helio}$\tablenotemark{b}} &  
\colhead{W} \\ 
\multicolumn{1}{l}{} &  
\colhead{({\AA})} &  
\colhead{(km/s)} & 
\colhead{(m\,\AA)}   
} 
 
\startdata  
\ion{Ca}{2}\,K     & 3932.55 & $-$86.3 & 128.5 \\ 
\ion{Na}{1}\,D2, 1 & 5888.31 & $-$83.6 & \nodata \\ 
\ion{Na}{1}\,D1, 1 & 5894.29 & $-$83.1 & \nodata \\ 
\ion{Na}{1}\,D2, 2 & 5888.14 & $-$92.2 & \nodata \\ 
\ion{Na}{1}\,D1, 2 & 5894.1: & $-$91.8:& \nodata \\  
\ion{Na}{1}\,D2, 1+2 & \nodata & \nodata & $>41.8$ \\ 
\ion{Na}{1}\,D1, 1+2 & \nodata & \nodata & $>12.2$  
\enddata 
\tablenotetext{a}{Wavelength is measured on the laboratory scale of the star.} 
\tablenotetext{b}{For Ca and Na, the velocity is estimated relative to
  each stellar line. } 
\end{deluxetable}  
 
\clearpage 
\begin{deluxetable}{lrrrr} 
\tabletypesize{\tiny}
\tablecolumns{5} 
\tablewidth{0pc} 
\tablecaption{\label{eqw} Equivalent width measurements} 
\tablehead{ 
\colhead{Ion}& 
\colhead{$\lambda$\,({\AA})}& 
\colhead{$\chi$\,(eV)} &  
\colhead{$\log\,gf$}& 
\colhead{$W$\,(m{\AA})}} 
\startdata 
Na  I & 5889.95 &  0.00 &     0.12 &   49.0 \\ 
Na  I & 5895.92 &  0.00 &  $-$0.18 &   30.1 \\ 
Mg  I & 3829.35 &  2.71 &  $-$0.21 &   83.8 \\ 
Mg  I & 3832.30 &  2.71 &     0.27 &  105.8 \\ 
Mg  I & 3838.29 &  2.71 &     0.49 &  117.4 \\ 
Mg  I & 4702.99 &  4.33 &  $-$0.38 &    9.4 \\ 
Mg  I & 5172.68 &  2.71 &  $-$0.40 &   84.3 \\ 
Mg  I & 5183.60 &  2.72 &  $-$0.18 &   97.3 \\ 
Al  I & 3961.53 &  0.01 &  $-$0.34 &   41.3 \\ 
Si  I & 3905.52 &  1.91 &  $-$1.09 &   99.6 \\ 
Ca  I & 4226.73 &  0.00 &     0.24 &   93.1 \\ 
Ca  I & 4318.65 &  1.89 &  $-$0.21 &    5.4 \\ 
Ca  I & 4434.96 &  1.89 &     0.00 &    9.3 \\  
Ca  I & 4435.69 &  1.89 &  $-$0.52 &    3.9 \\ 
Ca  I & 4454.78 &  1.90 &     0.26 &   16.2 \\ 
Ca II & 3181.27 &  3.15 &  $-$0.46 &   47.1 \\ 
Ca II & 3706.02 &  3.12 &  $-$0.48 &   49.4 \\ 
Ca II & 3736.90 &  3.15 &  $-$0.17 &   64.2 \\ 
Sc II & 3353.72 &  0.31 &     0.25 &   21.0 \\ 
Sc II & 3535.71 &  0.31 &  $-$0.47 &    4.6 \\ 
Sc II & 3572.53 &  0.02 &     0.27 &   40.8 \\ 
Sc II & 3576.34 &  0.01 &     0.01 &   27.4 \\ 
Sc II & 3580.93 &  0.00 &  $-$0.15 &   29.3 \\ 
Sc II & 3590.47 &  0.02 &  $-$0.55 &   16.8 \\ 
Sc II & 3630.74 &  0.01 &     0.22 &   39.2 \\ 
Sc II & 3642.78 &  0.00 &     0.13 &   43.6 \\ 
Sc II & 3645.31 &  0.02 &  $-$0.42 &   13.4 \\ 
Sc II & 3651.79 &  0.01 &  $-$0.53 &   11.2 \\ 
Sc II & 4246.82 &  0.31 &     0.24 &   30.3 \\ 
Sc II & 4415.56 &  0.60 &  $-$0.67 &    4.4 \\ 
Ti  I & 3653.49 &  0.05 &     0.22 &    8.6 \\ 
Ti  I & 3998.64 &  0.05 &     0.00 &    5.0 \\ 
Ti  I & 4533.25 &  0.85 &     0.53 &    6.5 \\ 
Ti  I & 4534.78 &  0.84 &     0.34 &    3.1 \\  
Ti II & 3148.05 &  0.00 &  $-$1.20 &   54.0 \\ 
Ti II & 3190.88 &  1.08 &     0.19 &   63.5 \\ 
Ti II & 3222.84 &  0.01 &  $-$0.48 &   69.0 \\ 
Ti II & 3229.20 &  0.00 &  $-$0.55 &   71.3 \\ 
Ti II & 3236.58 &  0.03 &     0.23 &   99.5 \\ 
Ti II & 3239.04 &  0.01 &     0.06 &   94.4 \\  
Ti II & 3241.99 &  0.00 &  $-$0.05 &   86.7 \\ 
Ti II & 3261.62 &  1.23 &     0.08 &   55.5 \\ 
Ti II & 3278.29 &  1.23 &  $-$0.21 &   45.6 \\ 
Ti II & 3282.32 &  1.22 &  $-$0.29 &   26.7 \\ 
Ti II & 3282.32 &  1.22 &  $-$0.29 &   26.7 \\ 
Ti II & 3287.66 &  1.89 &     0.34 &   40.4 \\ 
Ti II & 3302.11 &  0.15 &  $-$2.33 &    9.4 \\ 
Ti II & 3321.70 &  1.23 &  $-$0.32 &   37.7 \\ 
Ti II & 3322.94 &  0.15 &  $-$0.09 &   85.0 \\ 
Ti II & 3326.78 &  0.11 &  $-$1.18 &   55.5 \\ 
Ti II & 3329.45 &  0.14 &  $-$0.27 &   81.8 \\ 
Ti II & 3332.11 &  1.24 &  $-$0.15 &   45.7 \\ 
Ti II & 3335.20 &  0.12 &  $-$0.44 &   74.1 \\ 
Ti II & 3340.36 &  0.11 &  $-$0.61 &   62.8 \\ 
Ti II & 3343.76 &  0.15 &  $-$1.27 &   40.1 \\  
Ti II & 3349.04 &  0.61 &     0.47 &   78.1 \\ 
Ti II & 3361.22 &  0.03 &     0.28 &  102.8 \\ 
Ti II & 3372.80 &  0.01 &     0.27 &  100.6 \\ 
Ti II & 3380.28 &  0.05 &  $-$0.57 &   65.7 \\ 
Ti II & 3383.77 &  0.00 &     0.14 &   98.9 \\ 
Ti II & 3387.85 &  0.03 &  $-$0.43 &   76.4 \\ 
Ti II & 3394.58 &  0.01 &  $-$0.54 &   77.4 \\ 
Ti II & 3456.39 &  2.06 &  $-$0.23 &   11.7 \\ 
Ti II & 3477.19 &  0.12 &  $-$0.97 &   51.5 \\ 
Ti II & 3489.74 &  0.14 &  $-$1.92 &   16.5 \\ 
Ti II & 3491.07 &  0.11 &  $-$1.06 &   59.4 \\  
Ti II & 3573.73 &  0.57 &  $-$1.50 &   14.2 \\ 
Ti II & 3641.33 &  1.24 &  $-$0.71 &   14.1 \\ 
Ti II & 3685.19 &  0.57 &  $-$0.04 &   81.8 \\ 
Ti II & 3814.58 &  0.57 &  $-$1.70 &   14.3 \\ 
Ti II & 3913.48 &  1.12 &  $-$0.53 &   25.3 \\ 
Ti II & 4450.50 &  1.08 &  $-$1.51 &    6.3 \\ 
Ti II & 4533.97 &  1.24 &  $-$0.77 &   20.8 \\ 
Ti II & 4589.96 &  1.24 &  $-$1.62 &    3.8 \\ 
 
Cr  I & 3578.68 &  0.00 &     0.41 &   39.1 \\ 
Cr  I & 3593.48 &  0.00 &     0.31 &   44.4 \\ 
Cr  I & 4254.33 &  0.00 &  $-$0.11 &   21.7 \\ 
Cr  I & 4274.80 &  0.00 &  $-$0.23 &   16.7 \\ 
Cr  I & 4289.72 &  0.00 &  $-$0.36 &   20.4 \\ 
Cr II & 3124.97 &  2.45 &     0.30 &   37.6 \\ 
Cr II & 3408.76 &  2.48 &  $-$0.04 &   22.2 \\ 
Fe  I & 3225.79 &  2.40 &     0.38 &   43.4 \\ 
Fe  I & 3286.75 &  2.18 &  $-$0.17 &   21.7 \\ 
Fe  I & 3407.46 &  2.18 &  $-$0.02 &   30.5 \\ 
Fe  I & 3440.61 &  0.00 &  $-$0.67 &   93.9 \\ 
Fe  I & 3440.99 &  0.05 &  $-$0.96 &   86.2 \\ 
Fe  I & 3443.88 &  0.09 &  $-$1.37 &   69.4 \\ 
Fe  I & 3475.45 &  0.09 &  $-$1.05 &   73.9 \\ 
Fe  I & 3476.70 &  0.12 &  $-$1.51 &   68.5 \\ 
Fe  I & 3490.57 &  0.05 &  $-$1.11 &   87.8 \\ 
Fe  I & 3497.84 &  0.11 &  $-$1.55 &   60.9 \\ 
Fe  I & 3521.26 &  0.92 &  $-$0.99 &   41.4 \\ 
Fe  I & 3536.56 &  2.88 &     0.12 &    9.7 \\ 
Fe  I & 3554.93 &  2.83 &     0.54 &   28.1 \\ 
Fe  I & 3558.51 &  0.99 &  $-$0.63 &   54.7 \\ 
Fe  I & 3565.38 &  0.96 &  $-$0.13 &   80.3 \\ 
Fe  I & 3570.10 &  0.92 &     0.15 &   91.4 \\ 
Fe  I & 3581.19 &  0.86 &     0.41 &  102.0 \\  
Fe  I & 3585.32 &  0.96 &  $-$0.80 &   64.0 \\ 
Fe  I & 3585.71 &  0.92 &  $-$1.19 &   36.1 \\ 
Fe  I & 3586.11 &  3.24 &     0.17 &   12.1 \\ 
Fe  I & 3586.99 &  0.99 &  $-$0.80 &   49.5 \\ 
Fe  I & 3603.20 &  2.69 &  $-$0.26 &   15.4 \\ 
Fe  I & 3608.86 &  1.01 &  $-$0.10 &   73.8 \\ 
Fe  I & 3618.77 &  0.99 &     0.00 &   75.8 \\ 
Fe  I & 3621.46 &  2.73 &  $-$0.02 &   15.6 \\ 
Fe  I & 3622.00 &  2.76 &  $-$0.15 &    7.9 \\ 
Fe  I & 3631.46 &  0.96 &  $-$0.04 &   81.6 \\ 
Fe  I & 3640.39 &  2.73 &  $-$0.11 &   14.5 \\ 
Fe  I & 3647.84 &  0.92 &  $-$0.19 &   77.9 \\ 
Fe  I & 3687.46 &  0.86 &  $-$0.83 &   61.8 \\ 
Fe  I & 3689.46 &  2.94 &  $-$0.17 &    7.5 \\ 
Fe  I & 3709.25 &  0.92 &  $-$0.65 &   62.1 \\  
Fe  I & 3719.94 &  0.00 &  $-$0.43 &  109.1 \\ 
Fe  I & 3727.62 &  0.96 &  $-$0.63 &   65.2 \\  
Fe  I & 3737.13 &  0.05 &  $-$0.57 &   93.7 \\ 
Fe  I & 3743.36 &  0.99 &  $-$0.79 &   63.9 \\ 
Fe  I & 3786.68 &  1.01 &  $-$2.22 &    8.3 \\  
Fe  I & 3787.88 &  1.01 &  $-$0.86 &   53.1 \\ 
Fe  I & 3790.09 &  0.99 &  $-$1.76 &   11.4 \\ 
Fe  I & 3799.55 &  0.96 &  $-$0.85 &   57.7 \\  
Fe  I & 3812.96 &  0.96 &  $-$1.05 &   53.5 \\ 
Fe  I & 3815.84 &  1.49 &     0.24 &   70.2 \\ 
Fe  I & 3820.43 &  0.86 &     0.12 &  120.2 \\ 
Fe  I & 3824.44 &  0.00 &  $-$1.36 &   83.6 \\ 
Fe  I & 3825.88 &  0.92 &  $-$0.04 &   92.2 \\ 
Fe  I & 3827.82 &  1.56 &     0.06 &   64.3 \\ 
Fe  I & 3834.22 &  0.96 &  $-$0.30 &   76.0 \\ 
Fe  I & 3840.44 &  0.99 &  $-$0.51 &   66.4 \\ 
Fe  I & 3841.05 &  1.61 &  $-$0.05 &   58.7 \\ 
Fe  I & 3843.26 &  3.05 &  $-$0.24 &    6.2 \\  
Fe  I & 3849.97 &  1.01 &  $-$0.87 &   55.4 \\ 
Fe  I & 3850.82 &  0.99 &  $-$1.73 &   22.5 \\  
Fe  I & 3856.37 &  0.05 &  $-$1.29 &   77.6 \\ 
Fe  I & 3859.91 &  0.00 &  $-$0.71 &  104.8 \\ 
Fe  I & 3865.52 &  1.01 &  $-$0.95 &   50.1 \\ 
Fe  I & 3878.02 &  0.96 &  $-$0.90 &   53.9 \\ 
Fe  I & 3887.05 &  0.92 &  $-$1.14 &   52.0 \\  
Fe  I & 3902.95 &  1.56 &  $-$0.47 &   52.0 \\ 
Fe  I & 3906.48 &  0.11 &  $-$2.24 &   41.6 \\  
Fe  I & 3917.18 &  0.99 &  $-$2.15 &   13.1 \\ 
Fe  I & 3920.26 &  0.12 &  $-$1.75 &   66.4 \\ 
Fe  I & 3922.91 &  0.05 &  $-$1.65 &   74.9 \\  
Fe  I & 3930.30 &  0.09 &  $-$1.49 &   74.5 \\ 
Fe  I & 3956.68 &  2.69 &  $-$0.43 &    5.4 \\ 
Fe  I & 4005.24 &  1.56 &  $-$0.58 &   40.0 \\ 
Fe  I & 4045.81 &  1.49 &     0.28 &   76.5 \\ 
Fe  I & 4063.59 &  1.56 &     0.06 &   67.3 \\  
Fe  I & 4071.74 &  1.61 &  $-$0.02 &   60.7 \\ 
Fe  I & 4132.06 &  1.61 &  $-$0.68 &   36.4 \\ 
Fe  I & 4134.68 &  2.83 &  $-$0.65 &    4.4 \\ 
Fe  I & 4143.87 &  1.56 &  $-$0.51 &   41.7 \\ 
%
Fe  I & 4187.04 &  2.45 &  $-$0.51 &    8.9 \\ 
Fe  I & 4199.10 &  3.05 &     0.16 &   13.0 \\ 
Fe  I & 4202.03 &  1.49 &  $-$0.69 &   40.6 \\ 
Fe  I & 4222.21 &  2.45 &  $-$0.91 &    6.9 \\ 
Fe  I & 4227.43 &  3.33 &     0.27 &    6.8 \\ 
Fe  I & 4233.60 &  2.48 &  $-$0.58 &   10.6 \\ 
Fe  I & 4250.12 &  2.47 &  $-$0.38 &    7.8 \\ 
Fe  I & 4250.79 &  1.56 &  $-$0.71 &   35.3 \\ 
Fe  I & 4260.47 &  2.40 &     0.08 &   37.3 \\ 
Fe  I & 4271.76 &  1.49 &  $-$0.17 &   61.9 \\ 
Fe  I & 4325.76 &  1.61 &     0.01 &   64.0 \\ 
Fe  I & 4375.93 &  0.00 &  $-$3.03 &   20.3 \\ 
Fe  I & 4383.54 &  1.49 &     0.21 &   81.0 \\ 
Fe  I & 4404.75 &  1.56 &  $-$0.15 &   63.1 \\ 
Fe  I & 4415.12 &  1.61 &  $-$0.62 &   42.2 \\ 
Fe  I & 4427.31 &  0.05 &  $-$2.92 &   15.8 \\ 
Fe  I & 4430.61 &  2.22 &  $-$1.73 &    2.5 \\ 
Fe  I & 4447.72 &  2.22 &  $-$1.34 &    5.5 \\ 
Fe  I & 4459.12 &  2.18 &  $-$1.28 &    5.6 \\ 
Fe  I & 4461.65 &  0.09 &  $-$3.21 &    9.1 \\ 
Fe  I & 4494.56 &  2.20 &  $-$1.14 &    5.7 \\ 
Fe  I & 4528.61 &  2.18 &  $-$0.89 &   10.0 \\  
Fe  I & 4602.94 &  1.49 &  $-$2.21 &    4.6 \\ 
Fe  I & 4891.49 &  2.85 &  $-$0.11 &    8.1 \\ 
Fe  I & 4918.99 &  2.85 &  $-$0.34 &   12.4 \\ 
Fe  I & 4920.50 &  2.83 &     0.07 &   14.6 \\  
Fe  I & 5269.54 &  0.86 &  $-$1.32 &   55.2 \\ 
Fe II & 3295.82 &  1.08 &  $-$2.90 &   12.7 \\ 
Fe II & 4233.17 &  2.58 &  $-$1.81 &    5.7 \\ 
Fe II & 4522.63 &  2.84 &  $-$2.03 &    2.8 \\ 
Fe II & 4583.84 &  2.81 &  $-$1.74 &    3.8 \\ 
Fe II & 4923.93 &  2.89 &  $-$1.21 &   12.2 \\  
Fe II & 5018.44 &  2.89 &  $-$1.22 &   18.1 \\ 
Fe II & 5169.03 &  2.89 &  $-$1.30 &   18.8 \\ 
Co  I & 3409.17 &  0.51 &  $-$0.23 &   24.1 \\ 
Co  I & 3412.33 &  0.51 &     0.03 &   29.2 \\ 
Co  I & 3412.63 &  0.00 &  $-$0.78 &   30.3 \\ 
Co  I & 3443.64 &  0.51 &  $-$0.01 &   24.7 \\ 
Co  I & 3449.44 &  0.43 &  $-$0.50 &   29.8 \\  
Co  I & 3483.41 &  0.51 &  $-$1.00 &   13.2 \\ 
Co  I & 3502.28 &  0.43 &     0.07 &   38.0 \\ 
Co  I & 3502.62 &  0.17 &  $-$1.24 &    9.1 \\ 
Co  I & 3518.35 &  1.05 &     0.07 &   18.9 \\ 
Co  I & 3521.57 &  0.43 &  $-$0.58 &   16.7 \\  
Co  I & 3523.43 &  0.63 &  $-$0.44 &   25.9 \\ 
Co  I & 3526.84 &  0.00 &  $-$0.62 &   21.9 \\ 
Co  I & 3529.02 &  0.17 &  $-$0.88 &   24.5 \\ 
Co  I & 3529.81 &  0.51 &  $-$0.07 &   23.8 \\ 
Co  I & 3842.05 &  0.92 &  $-$0.77 &    2.5 \\ 
Co  I & 4121.31 &  0.92 &  $-$0.30 &    7.6 \\ 
Ni  I & 3232.93 &  0.00 &  $-$1.01 &   49.0 \\  
Ni  I & 3243.05 &  0.03 &  $-$1.30 &   43.6 \\ 
Ni  I & 3320.25 &  0.17 &  $-$1.42 &   33.7 \\ 
Ni  I & 3369.56 &  0.00 &  $-$0.66 &   63.6 \\ 
Ni  I & 3374.21 &  0.03 &  $-$1.76 &   28.8 \\ 
Ni  I & 3391.04 &  0.00 &  $-$1.05 &   52.4 \\ 
Ni  I & 3392.98 &  0.03 &  $-$0.54 &   65.5 \\ 
Ni  I & 3413.93 &  0.11 &  $-$1.72 &   25.2 \\ 
Ni  I & 3423.70 &  0.21 &  $-$0.76 &   64.7 \\ 
Ni  I & 3433.55 &  0.03 &  $-$0.67 &   59.0 \\  
Ni  I & 3437.27 &  0.00 &  $-$1.19 &   55.7 \\ 
Ni  I & 3452.89 &  0.11 &  $-$0.91 &   58.6 \\ 
Ni  I & 3461.65 &  0.03 &  $-$0.35 &   73.8 \\ 
Ni  I & 3469.48 &  0.28 &  $-$1.82 &   18.7 \\ 
Ni  I & 3472.54 &  0.11 &  $-$0.81 &   56.8 \\ 
Ni  I & 3483.77 &  0.28 &  $-$1.11 &   50.2 \\ 
Ni  I & 3492.95 &  0.11 &  $-$0.25 &   71.2 \\ 
Ni  I & 3500.85 &  0.17 &  $-$1.28 &   33.0 \\ 
Ni  I & 3519.76 &  0.28 &  $-$1.41 &   33.5 \\ 
Ni  I & 3524.53 &  0.03 &     0.01 &   87.3 \\ 
Ni  I & 3566.37 &  0.42 &  $-$0.24 &   60.0 \\ 
Ni  I & 3571.86 &  0.17 &  $-$1.14 &   51.2 \\ 
Ni  I & 3597.70 &  0.21 &  $-$1.10 &   56.3 \\  
Ni  I & 3807.14 &  0.42 &  $-$1.21 &   26.7 \\ 
Ni  I & 3858.29 &  0.42 &  $-$0.95 &   40.5  
\enddata 
\end{deluxetable} 
\clearpage

\begin{deluxetable}{lccc} 
\tablecolumns{4} 
\tablewidth{0pc} 
\tablecaption{\label{stell_par} Stellar parameters of HE~1300$+$0157} 
 
\tablehead{ 
\colhead{}&\multicolumn{2}{c}{This work}&\colhead{\citet{heresII}}\\ 
\colhead{Parameter}&\colhead{Subgiant}&\colhead{Dwarf\tablenotemark{a}}&\colhead{Subgiant} 
} 
\startdata 
\mbox{$T_{\rm eff}$} [K]&$5450\pm70$ &$5450\pm70$ &$5411 \pm100$  \\ 
$\log g$\,(cgs)         &$3.2 \pm0.3$&$4.6 \pm0.5$&$3.38 \pm0.44$ \\ 
$\mbox{[Fe/H]}$         &$-3.88\pm0.2$&$-3.31\pm0.2$&$-3.76\pm0.19$ \\ 
$v_{\rm micr}$ [km\,s$^{-1}$]&$1.54\pm0.3$&$0.80\tablenotemark{b}\pm0.3$&$1.43\pm0.27$ 
\enddata 
\tablenotetext{a}{The dwarf case stellar parameters are listed for comparison 
purposes only.} 
\tablenotetext{b}{Derived only from Ti lines.} 
\end{deluxetable}

\begin{deluxetable}{lrrcr} 
\tablecolumns{5} 
\tablewidth{0pc} 
\tablecaption{\label{balmer_tab} Balmer jump analysis data} 
\tablehead{ 
\colhead{Star}& 
\colhead{B.~J. mag}& 
\colhead{$(V-I)_{0}$} & 
\colhead{Type}& 
\colhead{Ref.} 
} 
\startdata 
HD~134439              & 0.03  & 0.89 & dwarf    & 1 \\ 
BD~+$66^{\circ}$ 0268 & 0.05  & 0.83 & dwarf    & 1, 2 \\ 
G~188-30               & 0.06  & 0.84 & dwarf    & 1 \\ 
HD~188510              & 0.08  & 0.75 & dwarf    & 1, 2 \\ 
CD~$-61^{\circ}$ 282  & 0.18  & 0.66 & dwarf    & 5, 6 \\ 
CD~$-31^{\circ}$ 622  & 0.21  & 0.85 & subgiant & 3, 4 \\ 
HD~45282               & 0.21  & 0.79 & subgiant & 1, 4 \\ 
HD~161770              & 0.22  & 0.81 & subgiant & 1, 4 \\ 
BD~+$37^{\circ}$ 1458 & 0.23  & 0.78 & subgiant & 1, 2 \\ 
HE~1300+0157          & 0.23  & 0.80 & subgiant &  \\ 
HD~140283              & 0.26  & 0.65 & subgiant & 1, 6 \\ 
HD~132475              & 0.27  & 0.64 & subgiant & 1, 6 \\ 
HD~160617              & 0.35  & 0.57 & subgiant & 1, 6 	 
\enddata  
 
\tablecomments{Refs. (1)-- \citet{schuster88}, (2) -- \citet{lairdpm88}, (3) 
  -- \citet{eggen90}, (4) -- \citet{Anthony-Twarog_by94}, (5) -- 
    \citet{nissen1997}, (6) -- \citet{schuster89}}  
 
\end{deluxetable}

\clearpage 
\begin{deluxetable}{lcrrrrrrrrrc} 
\tabletypesize{\scriptsize}
\tablecaption{\label{abund} LTE abundances of HE~1300$+$0157 }   
\tablewidth{0pt}  
\tablehead{  
\colhead{}&\colhead{}&\colhead{}&\colhead{}& 
\multicolumn{3}{c}{Subgiant}& \colhead{}& \multicolumn{3}{c}{Dwarf}& 
\colhead{}\\ 
\cline{5-7} 
\cline{9-11}\\ 
\colhead{Element} &  
\colhead{Ion} &  
\colhead{$\log\epsilon$\,(X)$_{\odot}$}&  
\colhead{3D/NLTE\tablenotemark{a}} & 
\colhead{$\log\epsilon$\,(X)} & 
\colhead{$\mbox{[X/H]}$}&  
\colhead{$\mbox{[X/Fe]}$}&  
\colhead{}& 
\colhead{$\log\epsilon$\,(X)} & 
\colhead{$\mbox{[X/H]}$}&  
\colhead{$\mbox{[X/Fe]}$}&  
\colhead{$N_{\rm lines}\tablenotemark{b}$}  } 
\startdata  
C &CH&8.39&\nodata&   5.89& $-2.50$&   1.38&&   5.48& $-2.91$&   0.40& syn\\ 
  &  &    &$-0.60$&   5.29& $-3.10$&   0.78&&   4.88& $-3.51$&$-$0.20& syn\\ 
N &NH&7.78&\nodata&$<5.12$&$<-2.66$&$<1.22$&&$<4.70$&$<-3.08$&$<0.23$&syn\\ 
  &  &    &$-0.60$&$<4.52$&$<-3.26$&$<0.62$&&$<4.10$&$<-3.68$&$<-0.37$&syn\\ 
C+N&  &    &\nodata&\nodata&$<-2.53$&$<1.35$&&\nodata&$<-2.94$&$<0.37$&   \\ 
   &  &    &$-0.60$&\nodata&$<-3.13$&$<0.75$&&\nodata&$<-3.54$&$<-0.23$&   \\ 
O &OH&8.66&\nodata& $6.54$& $-2.12$& $1.76$&& $6.13$& $-2.53$& $0.78$& syn\\ 
  &  &    &$-0.60$& $5.94$& $-2.72$& $1.16$&& $5.53$& $-3.13$& $0.18$& syn\\ 
Li& 1&1.05&\nodata&   1.06&\nodata&\nodata&&   1.12&\nodata&\nodata& syn \\ 
Na& 1&6.17&\nodata&   2.44&$-$3.73&   0.15&&   2.46&$-$3.71&$-$0.40& 2 \\  
  &  &    &$-$0.40&   2.04&$-$4.13&$-$0.25&&   2.06&$-$4.11&$-$0.80& 2 \\  
Mg& 1&7.53&\nodata&   4.10&$-$3.43&   0.45&&   3.89&$-$3.64&$-$0.33& 6 \\  
  &  &    &  +0.15&   4.25&$-$3.28&   0.60&&   4.04&$-$3.49&$-$0.18& 6 \\  
Al& 1&6.37&\nodata&   2.04&$-$4.33&$-$0.45&&   2.08&$-$4.29&$-$0.98& 1 \\  
  &  &    &  +0.50&   2.54&$-$3.83&$-$0.05&&   2.58&$-$3.79&$-$0.48& 1 \\ 
Si& 1&7.51&\nodata&   4.50&$-$3.01&   0.87&&   4.29&$-$3.22&   0.09& 1 \\  
Ca& 1&6.31&\nodata&   2.98&$-$3.33&   0.55&&   2.99&$-$3.32&$-$0.01& 5 \\  
Ca& 2&6.31&\nodata&   2.75&$-$3.56&   0.32&&   3.16&$-$3.15&   0.16& 4 \\  
Sc& 2&3.05&\nodata&$-$0.37&$-$3.42&   0.46&&   0.18&$-$2.87&   0.44& 12\\  
Ti& 1&4.90&\nodata&   1.66&$-$3.24&   0.64&&   1.72&$-$3.18&   0.13& 4 \\  
Ti& 2&4.90&\nodata&   1.77&$-$3.13&   0.75&&   2.31&$-$2.59&   0.72& 40\\  
Cr& 1&5.64&\nodata&   1.51&$-$4.13&$-$0.25&&   1.60&$-$4.04&$-$0.73& 5 \\  
Cr& 2&5.64&\nodata&   1.65&$-$3.99&$-$0.11&&   2.25&$-$3.39&$-$0.08& 2 \\  
Mn&1 &5.39&\nodata&   1.05&$-$4.34&$-$0.46&&   1.10&$-$4.29&$-$0.52& 1 \\  
Mn& 2&5.39&\nodata&   0.89&$-$4.50&$-$0.62&&   1.44&$-$3.95&$-$0.98& 1 \\  
Fe& 1&7.45&\nodata&   3.72&$-$3.73&   0.15&&   3.75&$-$3.70&$-$0.39& 101\\  
  &  &    &  +0.20&   3.92&$-$3.53&   0.35&&   3.95&$-$3.50&$-$0.19& 101\\  
Fe& 2&7.45&\nodata&   3.57&$-$3.88&   0.00&&   4.14&$-$3.31&   0.00& 7 \\  
Co& 1&4.92&\nodata&   1.80&$-$3.12&   0.76&&   1.92&$-$3.00&   0.31& 16\\  
Ni& 1&6.23&\nodata&   2.61&$-$3.62&   0.26&&   2.74&$-$3.49&$-$0.18& 25\\\tableline\\ 
V & 1&4.00&\nodata& $<1.15$&$<-2.85$& $<1.03$&& $<1.21$&$<-2.79$& $<0.52$&4379\\ 
Zn& 1&4.60&\nodata& $<1.63$&$<-2.97$& $<0.91$&& $<1.99$&$<-2.61$& $<0.70$&4810\\ 
Sr& 2&2.92&\nodata&$<-2.64$&$<-5.56$&$<-1.68$&&$<-2.11$&$<-5.03$&$<-1.72$&4077\\ 
  &  &    &   0.30&$<-2.34$&$<-5.26$&$<-1.38$&&$<-1.81$&$<-4.73$&$<-1.42$&4077\\ 
Y & 2&2.21&\nodata&$<-0.93$&$<-3.14$& $<0.74$&&$<-0.41$&$<-2.62$& $<0.69$&4884\\  
Zr& 2&2.59&\nodata&$<-0.07$&$<-2.66$& $<1.22$&& $<0.45$&$<-2.14$& $<1.17$&4497\\  
Ba& 2&2.17&\nodata&$<-2.56$&$<-4.73$&$<-0.85$&&$<-2.04$&$<-4.21$&$<-0.90$&4554\\ 
  &  &    &  +0.15&$<-2.41$&$<-4.58$&$<-0.70$&&$<-1.89$&$<-4.06$&$<-0.75$&4554\\ 
Eu& 2&0.52&\nodata&$<-1.80$&$<-2.32$& $<1.56$&&$<-1.27$&$<-1.79$& $<1.52$&4129\\ 
Pb& 1&2.00&\nodata&$< 0.90$&$<-1.10$& $<2.78$&&$< 0.98$&$<-1.02$& $<2.29$&4057 
 
\enddata              
\tablenotetext{a}{Where indicated, the correction has been added to the LTE 
  abundance.} 
\tablenotetext{b}{``Syn'' indicates the use of spectrum synthesis for the 
  abundance determination. For upper limits the wavelength of the line used is 
  given.} 
\end{deluxetable}

 
\clearpage 
\begin{deluxetable}{lrrrrrr} 
\tablecolumns{3} 
\tablewidth{0pc} 
\tablecaption{\label{err} Abundance Uncertainties} 
\tablehead{\colhead{Element}&\colhead{Ion}&\colhead{Random}& 
\colhead{$\Delta$\mbox{T$_{\rm eff}$}}&\colhead{$\Delta\log g$}& 
\colhead{$\Delta v_{micr}$}&\colhead{Root Mean}\\ 
\colhead{}&\colhead{}&\colhead{error}&\colhead{+100\,K}& 
\colhead{$+$0.3\,dex}&\colhead{+0.3\,km\,s$^{-1}$}&\colhead{Square}} 
\startdata 
C &CH& 0.15& 0.19 &  0.11 &    0.00 & 0.27 \\ 
O &OH& 0.30& 0.22 &  0.11 &    0.00 & 0.39 \\ 
Li& 1& 0.05& 0.08 &  0.01 &    0.00 & 0.09 \\ 
Na& 1& 0.06& 0.09 &$-$0.01& $-$0.02 & 0.10 \\ 
Mg& 1& 0.07& 0.09 &$-$0.06&    0.06 & 0.14 \\ 
Al& 1& 0.06& 0.10 &  0.00 & $-$0.03 & 0.12 \\ 
Si& 1& 0.06& 0.12 &$-$0.07& $-$0.10 & 0.18 \\ 
Ca& 1& 0.08& 0.08 &$-$0.01& $-$0.03 & 0.12 \\ 
Ca& 2& 0.06& 0.05 &  0.10 & $-$0.05 & 0.14 \\ 
Sc& 2& 0.07& 0.07 &  0.11 & $-$0.02 & 0.15 \\ 
Ti& 1& 0.10& 0.10 &$-$0.01& $-$0.01 & 0.14 \\ 
Ti& 2& 0.06& 0.04 &  0.08 & $-$0.12 & 0.16 \\ 
Cr& 1& 0.08& 0.12 &$-$0.01& $-$0.03 & 0.15 \\ 
Cr& 2& 0.07& 0.04 &  0.10 & $-$0.02 & 0.13 \\ 
Mn& 1& 0.06& 0.12 &  0.00 & $-$0.01 & 0.13 \\ 
Mn& 2& 0.06& 0.05 &  0.10 &    0.00 & 0.13 \\ 
Fe& 1& 0.06& 0.12 &$-$0.02& $-$0.08 & 0.16 \\ 
Fe& 2& 0.08& 0.02 &  0.11 & $-$0.01 & 0.14 \\ 
Co& 1& 0.08& 0.12 &  0.00 & $-$0.03 & 0.15 \\ 
Ni& 1& 0.06& 0.15 &$-$0.03& $-$0.11 & 0.20 \\

\enddata 
\end{deluxetable}

\clearpage 
\clearpage 
 
\clearpage 
 
\clearpage 
 
\clearpage 
\clearpage 
\clearpage 

\clearpage 
\clearpage 
\clearpage 

\clearpage 
\clearpage 
\clearpage 
 
\clearpage 
\clearpage 
\clearpage 
 
\end{document}